\documentclass[12pt]{article}
\usepackage[a4paper,margin=1in]{geometry}
\usepackage{amsmath}
\usepackage{graphicx}
\usepackage{natbib}
\usepackage{url} 
\usepackage{changepage}
\usepackage[utf8x]{inputenc}
\usepackage{textcomp,marvosym}
\usepackage{fixltx2e}
\usepackage{amsmath,amssymb}
\usepackage{nameref,hyperref}
\usepackage{authblk}
\usepackage{algorithm}
\usepackage{algpseudocode}
\usepackage{dsfont}
\usepackage{xcolor}
\usepackage{subcaption}
\usepackage{enumitem}
\usepackage{caption,setspace}
\usepackage{pdflscape}
\usepackage{graphicx}
\usepackage{placeins}
\usepackage{mathtools}
\usepackage[title]{appendix}

\renewcommand{\exp}[1]{\mathrm{exp}\!\left( #1 \right)}

\renewcommand{\mid}{\:\middle|\:}
\newcommand{\T}{^\mathrm{T}} 
\newcommand{\expect}[1]{\mathbb{E}\!\left[ #1 \right]} 
\newcommand{\condexpect}[2]{\mathbb{E}\!\left[ #1 \mid #2 \right]} 
\newcommand{\var}[1]{\mathrm{Var}[ #1 ]} 
\newcommand{\iid}{\overset{\mathrm{iid}}{\sim}}
\newcommand{\cond}[2]{\left( #1 \mid #2 \right)} 
\newcommand{\pdf}[1]{\pi\!\left( #1 \right)}
\newcommand{\condpdf}[2]{\pi\!\cond{#1}{#2}}
\newcommand{\approxcondpdf}[2]{\tilde{\pi}\!\cond{#1}{#2}}
\newcommand{\condprob}[2]{\mathrm{Pr}\!\cond{#1}{#2} } 
\newcommand{\rmd}{\,\mathrm{d}} 
\newcommand{\dydx}[2]{\dfrac{\rmd #1}{\rmd #2}} 

\newcommand{\data}[1]{y_{\mathrm{#1}}}

\newcommand{\approxdata}[1]{\tilde{y}_{\mathrm{#1}}}

\newcommand{\likelihood}[2]{\pi\!\left( #1 \mid #2 \right)} 

\newcommand{\indd}[2]{\mathds{1}_{#1}\left(#2\right)}

\renewcommand{\eqref}[1]{Equation~(\ref{#1})}

\begin{document}
	
\title{Simulation and inference methods for non-Markovian stochastic reaction networks}

\author[1]{Thomas~P. Steele}
\author[1,2,3]{David~J. Warne\footnote{To whom correspondence should be addressed. E-mail: david.warne@qut.edu.au}}

\affil[1]{School of Mathematical Sciences, Queensland University of Technology, Brisbane, Queensland 4001, Australia}
\affil[2]{Centre for Data Science, Queensland University of Technology, Brisbane, Queensland 4001, Australia}
\affil[3]{ARC Centre of Excellence for Mathematical Analysis of Cellular Systems (MACSYS), Queensland University of Technology, Brisbane, Queensland 4001, Australia}

\maketitle



%
\begin{abstract}
	Stochastic models of reaction networks are widely used to capture intrinsic noise in complex systems in the life sciences. Typical formulations of these models are based on Markov processes for which there is extensive research on efficient simulation and inference. However, there are complex processes in biology, such as gene transcription and translation, that introduce history dependent dynamics requiring non-Markovian processes to accurately capture the stochastic dynamics of the system. This greater realism comes with additional computational challenges for simulation and parameter inference. We develop efficient stochastic simulation algorithms for well-mixed non-Markovian stochastic reaction networks with stochastic delays that depend on system state and time. Our methods generalize the next reaction method and $\tau$-leaping method to support arbitrary inter-event time distributions while preserving computational scalability. We also introduce a coupling scheme to generate exact non-Markovian sample paths that are positively correlated to an approximate non-Markovian $\tau$-leaping sample path. This enables substantial computational gains for simulation and Bayesian inference through multilevel Monte Carlo and multifidelity schemes. We demonstrate the effectiveness of our approach using several non-Markovian examples, showing substantial gains in both simulation accuracy and inference efficiency. These results extend the practical applicability of non-Markovian models in systems biology and beyond.
\end{abstract} 


\section{Introduction}
\label{sec: 1 intro}

Stochastic modelling of system dynamics is an essential technique in many fields of science to describe processes driven by intrinsic noise~\cite{Abkowitz1996,Cox1965,Kaern2005}. Many real systems are well described by Markovian stochastic processes, where future evolution depends only on the current state and is conditionally independent of the past~\cite{Gillespie1976,Gillespie1977,Gillespie2001}. However, in many systems, history plays an essential role in shaping future dynamics. In such cases, non-Markovian stochastic processes are necessary for accurate simulation, prediction, and parameter inference~\cite{ Boguna2014,Browning2021,Cai2007,Fu2022,Whitaker2025,Yang2022}.

A particularly important application domain for stochastic models is the study of intracellular processes. Processes such as gene regulation are inherently stochastic, often due to the low copy numbers of key bio-molecules involved~\cite{Arkin1998,Elowitz2002,Gillespie1977}. The dynamics of such systems are governed by the chemical master equation (CME), which is analytically intractable in all but the simplest cases, necessitating computational approaches~\cite{Anderson2007}. Two primary problems for these systems are: the forward problem, that is, the stochastic simulation of a given system; and the inverse problem, that is, inferring model parameters given experimental data~\cite{Higham2008,Schnoerr2017,Warne2019}. 

For the forward problem, a variety of stochastic simulation methods are available. These include exact methods, such as the Gillespie direct method~\cite{Gillespie1976} and the next reaction method (NRM)~\cite{Cao2004,Gibson2000}, and approximate methods, such as the $\tau$-leaping method~\cite{Gillespie2001}. All of these methods are based on models of chemically reacting systems that are continuous-time Markov jump processes where state transitions occur instantaneously at reaction events times~\cite{Gillespie1977,Anderson2007}. However, the assumption of instantaneous reaction events is not always appropriate, particularly for complex reactions that occur in gene transcription and translation. In these cases, it may be more appropriate to model complex reactions using delayed reactions, due to the non-negligible duration between initiation and completion of a reaction~\cite{Bratsun2005, Brett2013,Meeussen2024,Miekisz2011}. The presence of delay reactions in a system presents a computational challenge as it becomes a non-Markovian stochastic process~\cite{Boguna2014, Bodnar2011,Cai2007,Fu2022}.

A popular approach to the inverse problem is Bayesian inference that updates knowledge of model parameters, $\theta \in \Theta$, through combining prior knowledge of these parameters with observed data, $\data{obs}$. This proceeds through the application of Bayes' theorem,
\begin{equation}
	\condpdf{\theta}{\data{obs}} = \frac{\likelihood{\data{obs}}{\theta}\pdf{\theta}}{\pdf{\data{obs}}},
	\label{eq:bayes}
\end{equation}
where $\pdf{\theta}$ is the prior probability density that represents current knowledge, $\likelihood{\data{obs}}{\theta}$ is the likelihood of an observation given model parameters, and $\pdf{\data{obs}}$ is the evidence or marginal likelihood that acts as a normalisation constant. Through \eqref{eq:bayes} we obtain the posterior probability density, $\condpdf{\theta}{\data{obs}}$, that represents updated knowledge informed by the prior and the data. Since the evidence term is usually intractable, sampling schemes such as Markov chain Monte Carlo (MCMC)~\cite{Hastings1970,Metropolis1953} and sequential Monte Carlo (SMC)~\cite{DelMoral2006} are used to sample to posterior distribution. However, these standard tools are not directly applicable in the context of partially observed stochastic processes, since the likelihood will almost always be intractable due to dependence on the CME solution~\cite{Erban2009,Golightly2011,Toni2008,Warne2019}. This is only rendered more challenging in the non-Markovian setting as the equivalent CME is non-trivial~\cite{Kanazawa2024}. 

Due to the likelihood intractability, we rely on so-called likelihood-free inference methods, also known as simulation-based inference. These methods rely on large numbers of model simulations as a substitute for direct likelihood evaluation~\cite{Cranmer2020,Sisson2018,Sunnaaker2013}. These include traditional approaches, such as approximate Bayesian computation (ABC)~\cite{Beaumont2002,Blum2010,Frazier2018,Pritchard1999,Tavare1997} and Bayesian synthetic likelihood (BSL)~\cite{Frazier2022,Price2017}, and modern machine learning approaches, such as neural likelihood estimation~\cite{Papamakarios19a} and neural posterior estimation~\cite{Papamakarios2016,Papamakarios2021a}. While machine learning approaches have been shown to require fewer stochastic simulations~\cite{Lueckmann2021a}, challenges still remain in terms of accuracy guarantees~\cite{Frazier2024,Wang2024}. Regardless of the specific approach, all likelihood-free and simulation-base inference approaches are computationally intensive, especially for complex or high-dimensional models, as they require large numbers of model simulations to obtain accurate posterior estimates.


Recent work has focused on acceleration strategies that leverage model approximations. Two notable approaches are multilevel Monte Carlo (MLMC)~\cite{Giles2008,Hikida2025,Jasra2019,Warne2018} and multifidelity schemes~\cite{Krouglova2025,Prescott2020,Rhee2015}. These techniques reduce the computational cost by combining large numbers of computationally inexpensive low-fidelity approximate simulations with relatively fewer computationally expensive high-fidelity simulations. Accuracy is maintained through low variance bias corrections that rely on variance reduction due to coupling simulation pairs~\cite{Anderson2012,Giles2008,Prescott2020}. As a result, MLMC and multifidelity approaches can achieve substantial improvements in computational efficiency,  often by orders of magnitude, without sacrificing inference accuracy~\cite{Hikida2025,Jasra2019,Krouglova2025,Prescott2020,Prescott2024,Warne2018,Warne2022}. 

Despite these advances, such multifidelity methods have been almost exclusively applied to Markovian systems~\cite{Prescott2020,Prescott2024,Warne2022}. Utilising multifidelity methods for non-Markovian stochastic processes introduces major challenges. In particular, many simulation schemes in the literature cannot simulate non-Markovian systems accurately~\cite{Voliotis2016}, and those that do are often restricted to delay time distributions that depend on time only~\cite{Anderson2007,Boguna2014}. Moreover, the generation of coupled sample paths in a non-Markovian setting, as required for MLMC and multifidelity approaches, requires careful handling of delay structures to maintain consistency and avoid statistical inaccuracies~\cite{Boguna2014, Kanazawa2024}. In addition, the focus of the literature has been on exact simulation schemes ~\cite{Anderson2007,Boguna2014,Voliotis2016}. As a result, the necessary approximate simulation and coupling schemes required for MLMC and multifidelity schemes have not been explored in the non-Markovian setting.

In this work, we address this challenge and enable MLMC and multifidelity simulation and inference approaches for non-Markovian reaction networks. We present a novel exact stochastic simulation that leads to a natural approximate stochastic simulation and coupling scheme. We begin by presenting the general framework for simulating jump processes with non-exponential inter-event times, then introduce non-Markovian extensions to both the NRM and the $\tau$-leaping method. Finally, we develop a novel coupling scheme for generating positively correlated approximate and exact non-Markovian sample paths. This enables the application of MLMC and multifidelity methods for the forwards and inverse problems for non-Markovian systems. We demonstrate the efficacy of our approach through the implementation of multifidelity ABC~\cite{Prescott2020}, however, our simulation algorithms are applicable to any of the current MLMC or multifidelity methods that rely on coupling for accelerating simulation~\cite{Anderson2012,Giles2008} and inference~\cite{Hikida2025,Prescott2021,Prescott2024,Warne2022}. Together, these contributions provide a foundation for efficient simulation and inference for non-Markovian systems, addressing a critical barrier to the practical implementation of these models.

\section{Methods}
\label{sec: 2 methods}



In this section, we establish some fundamental background and mathematical foundations that form a basis of our methods and algorithms. We begin by reviewing the standard formulation of a reaction network~\cite{Erban2007,Schnoerr2017,Warne2019} and then examine how delay reactions alter the structure and dynamics of these systems~\cite{Bodnar2011,Boguna2014}. From this, we derive the inter-event time distribution in non-Markovian systems and present a generalised representation of a reaction network that accommodates non-exponential waiting times. We then review the ABC approach to likelihood-free inference~\cite{Beaumont2002,Blum2010,Sisson2018} and acceleration using multifidelity schemes~\cite{Prescott2020,Prescott2024}. Finally, we introduce our algorithmic contributions to enable the application of multifidelity approaches in the non-Markovian setting. Specifically, we develop a non-Markovian extension to the NRM, a non-Markovian $\tau$-leaping scheme, and a novel coupling scheme for generating correlated approximate and exact non-Markovian simulation pairs.

\subsection{Modelling stochastic reaction networks}
\label{subsec: 2.1 NM-BCRN}

A reaction network consists of $\mathcal{N}$ chemical species, $\mathcal{X}_1, \mathcal{X}_2,\ldots, \mathcal{X}_{\mathcal{N}}$, with system state, \\$\mathbf{X}(t) = [X_1(t),\ldots,X_\mathcal{N}(t)]\T$, where $X_i(t) \in \mathbb{N}$ is the copy number of the $i$th chemical species. Species interact via a network of $\mathcal{M}$ reaction channels,
\begin{align}
	\sum_{i=1}^{\mathcal{N}} \nu^-_{ij} \mathcal{X}_i \xrightarrow{k_j} \sum_{i=1}^{\mathcal{N}} \nu^+_{ij} \mathcal{X}_i, \quad j=1,\dots,\mathcal{M}. \label{eqn: change state}
\end{align}
where $k_j > 0$ is the kinetic rate parameter for reaction $j$, and $\nu^{-} \in \mathbb{N}^{\mathcal{N}\times \mathcal{M}}$ and $\nu^+ \in \mathbb{N}^{\mathcal{N}\times \mathcal{M}}$ denote the reactant and product stoichiometric matrices, respectively. In this paper, for a matrix A, we adopt the notation $a_{i,j}$ for element in row $i$ and column $j$, and $a_{*,j}$ as the $j$th column of A. Given a system state $\mathbf{X}(t)$ the probability that reaction $j$ occurs in the interval $[t,t+\Delta t)$ is $\lambda_j(\mathbf{X}(t))\Delta t + \mathcal{O}(\Delta t^2)$ where $\lambda_j(\mathbf{X}(t))$ is the propensity function for the $j$th reaction channel. These propensity functions typically have the standard combinatorial form based on the law of mass action,
\begin{align}
	\lambda_j(\mathbf{X}(t)) = k_j \prod_{i=1}^{\mathcal{N}}\frac{X_i(t)!}{(X_i(t) - \nu^-_{ij})!}. \label{eq:prop_I}
\end{align}
However, nonlinear propensities, such as those based on Hill functions are also routinely used. In both cases, the propensity is a function of the system state only, this means that propensities are constant between reaction events.  
When a reaction event from channel $j$ occurs, the system state is instantaneously updated $\mathbf{X}(t+\tau)=\mathbf{X}(t)+\nu_{*,j}$ where $\nu_{*,j} = \nu^+_{*,j} - \nu^-_{*,j}$, and $\tau$ is the inter-event time. In this setting, the number of reaction events over time for channel $j$ can be described by an inhomogeneous Poisson process, this leads to the so-called Kurtz random time-change representation of a reaction network~\cite{Kurtz1972},
\begin{align}
	\mathbf{X}(t+\tau) = \mathbf{X}(t) + \sum_{j=1}^\mathcal{M} Y_j\left(\int_0^\tau\lambda_j\left(\mathbf{X}(t+s)\right)\rmd s\right)\nu_{*,j} \label{eqn:Kurtz},
\end{align}
where $Y_j(\cdot)$ is a unit-time homogeneous Poisson process.
The state transition probability function related to this process, $p(\mathbf{x}_t,\mathbf{x}_s) = \condprob{\mathbf{X}(t) = \mathbf{x}_t}{\mathbf{X}(s) = \mathbf{x}_s}$ for $t > s$, can be obtained though the solution to the forwards Kolmogorov equation, which is known as the CME in the biochemistry literature~\cite{Erban2009,Gillespie1992,Schnoerr2017},
\begin{align}
	\dydx{p(\mathbf{x}_t,\mathbf{x}_s)}{t} = \sum_{j=1}^{\mathcal{M}}\lambda_j(\mathbf{x}_t-\nu_j) p(\mathbf{x}_t-\nu_j,\mathbf{x}_s) - p(\mathbf{x}_t,\mathbf{x}_s)\sum_{j=1}^{\mathcal{M}}\lambda_j(\mathbf{x}_t)
	\label{eq:cme}
\end{align}
Unfortunately, the CME (\eqref{eq:cme}) is intractable for most realistic networks~\cite{Erban2007,Jahnke2006,Warne2019}.

\subsubsection{Inclusion of delay reactions}
\label{subsec: 2.1.2 slow reactions}

In complex biochemical processes, it may not always be appropriate to treat all reaction events as instantaneous. For example, the transcription process from DNA to mRNA and subsequent translation into proteins are complex biochemical processes that occur over a non-negligible period of time~\cite{Bodnar2011,Boguna2014,Bratsun2005,Meeussen2024,Pelissier2026}. This type of reaction cannot be adequately described by Equations (\ref{eqn: change state})--(\ref{eq:cme}). Instead these reactions must be described as \textit{delay reactions} where the interval between reactant consumption and product formation is explicitly modeled.

Suppose that of our $\mathcal{M}$ reactions $\mathcal{R} = \{1,2,\ldots, \mathcal{M}\}$ there is a subset $\mathcal{R}_D \subseteq \mathcal{R}$ that are delayed reactions and the remaining set $\mathcal{R}_I = \mathcal{R} \backslash \mathcal{R}_D$ are instantaneous reaction set so that $\mathcal{R} = \mathcal{R}_D \cup \mathcal{R}_I$ and $\mathcal{R}_D \cap \mathcal{R}_I = \emptyset$. This leads to the system
\begin{equation}
	\begin{split}
		\sum_{i=1}^{\mathcal{N}} \nu^-_{ij} \mathcal{X}_i &\xrightarrow{k_j} \sum_{i=1}^{\mathcal{N}} \nu^+_{ij} \mathcal{X}_i, \quad j \in \mathcal{R}_I \\
		\sum_{i=1}^{\mathcal{N}}  \nu^-_{ij} \mathcal{X}_i  &\xRightarrow[k^*_j]{k_j} \sum_{i=1}^{\mathcal{N}}  \nu^+_{ij} \mathcal{X}_i , \quad j \in \mathcal{R}_D. 
	\end{split}
	\label{eq:delay_bcrn}
\end{equation}
For instantaneous reactions $j \in \mathcal{R}_I$ that are denoted by $\rightarrow$, the standard formulation for the propensities $\lambda_j(\mathbf{X}(t))$ holds with $k_j > 0$ as the kinetic rate parameter (\eqref{eq:prop_I}). If the $j$th reaction is in $\mathcal{R}_D$ then it is a delayed reaction, which is denoted by $\Rightarrow$. This delay reaction has an initiation and a completion event, the initiation event occur according to a Poisson process with propensity $\lambda_j(\mathbf{X}(t))$ with kinetic rate parameter $k_j > 0$ just as in the instantaneous case. However the completion event, conditional on initiation time, $t^*_0$, occurs with propensity $\lambda^*_j(\mathbf{X}(t),t^*)$ with rate kinetic parameter $k^*_j > 0$ and internal time $t^* = t - t^*_0$ that leads to a non-exponential waiting-time distribution. 

Since each delay reaction has its own internal time, a delay reaction proceeds in two phases. First the reaction is initiated: suppose the $k$th event initiation event that occurs at time $t+\tau$ where $\tau$ is the inter-event time, the state vector updated according to $\mathbf{X}(t+\tau) = \mathbf{X}(t) - \nu_{*,j}^-$, and a new\textit{ delayed chemical species} is created, $\mathcal{X}_j^{*,k}$, for the $k$th initiation of the $j$th reaction (which is a delay reaction since $j \in \mathcal{R}_D$). The $k$th initiation time is set, $t_0^{*,k} = t+\tau$, is set and a completion reaction is introduced, 
\begin{align*}
	\mathcal{X}^{*,k}_j \xrightarrow{k^*_j} \sum_{i=1}^{\mathcal{N}} \mathcal{X}_i \nu^+_{ij}. \label{eqn: delay change state two step}
\end{align*}
The associated reaction network proceeds with this new reaction included in the network. Suppose at time $t_1$ the completion event for this $k$th initiation of reaction $j$ occurs at time $\tau_1$, the $k$th delayed chemical species $\mathcal{X}_j^{*,k}$ is removed along with the associated reaction, and the state is updated $\mathbf{X}(t_1+\tau_1) = \mathbf{X}(t_1) + \nu_{*,j}^+$. As formulated here, the virtual delayed chemical species are not considered part of the system state, but simply represent the ongoing progression of the complex reaction from initiation to completion. However, this can also be modified by defining a copy number $X_j^{*,k}(t) = 1$, at the $k$th initiation event, and including this as part of the system state $\mathbf{X}(t)$. This would enable the presence of ongoing reactions to affect any propensity values in the system for any reaction $j \in \mathcal{R}$. We only consider the total number of ongoing reactions as part of the state vector, that is,
\begin{equation}
	X_j^*(t) = \sum_{k=1}^\infty X_j^{*,k}(t). \label{eq:delay_state}
\end{equation}  
This leads to the extended state vector including the copy numbers for the $\mathcal{N}$ chemical species and counts of ongoing delayed reaction, that is, $\mathbf{X}(t) = [X_1(t),X_2(t),\ldots,X_\mathcal{N}(t),X_{j_1}^*(t),X_{j_2}^*(t),\ldots, X_{j_{\mathcal{D}}}^*(t)]$, where $\mathcal{D}$ is the number of delayed reaction and $\mathcal{R}_D = \{j_1,j_2,\ldots,j_{\mathcal{D}}\}$. Thus, if we consider a system with $\mathcal{N}$ chemical species and $\mathcal{M}$ reactions, of which $\mathcal{D} \leq \mathcal{M}$ are delayed reactions, then we will have an effective state vector of dimension $\mathcal{N} + \mathcal{D}$.

This leads to the following extension to the Kurtz respresentation, 
\begin{equation}
	\begin{split}
		\mathbf{X}(t+\tau) &= \mathbf{X}(t) + \sum_{j\in \mathcal{R}_I}\underbrace{Y_j\left(\int_0^\tau \lambda_j(\mathbf{X}(t+s))\,\text{d}s\right)}_{\text{Instantaneous reactions}}\nu_{*,j} \\ &\quad+\sum_{j\in \mathcal{R}_D}\underbrace{Y_j\left(\int_0^\tau \lambda_j(\mathbf{X}(t+s))\,\text{d}s\right)}_{\text{Delay initiation reaction}}(\eta_j-\nu^-_{*,j}) \\
		&\quad+ \sum_{j\in \mathcal{R}_D}\sum_{k=1}^\infty \underbrace{\min{\left\{1,Y^k_j\left(\int_0^\tau \lambda^*_j(\mathbf{X}(t+s), t_j^k + s)\,\text{d}s\right)\right\}}}_{\text{Completion of $k$th ongoing delay reaction}}(\nu^+_{*,j}-\eta_j).   
	\end{split}
	\label{eq:kurtz_delay}
\end{equation}
Here, $Y_j(\cdot)$ for $j \in \mathcal{R}$, and $Y_j^k(\cdot)$ for $j \in \mathcal{R}_D$ are $k \in \mathbb{Z}^+$ and unit rate homogeneous Poisson processes. We have $t_j^k = t -t_{j,0}^{*,k}$ as the internal time for the $k$ delay in the $j$th channel, where sequence, $t_{j,0}^{*,1} < t_{j,0}^{*,2} < \cdots < t_{j,0}^{*,k} < \cdots$, corresponds to the initiation event times for delay as generated by the initiation processes $Y_j(\cdot)$ for $j \in \mathcal{R}_D$. The vector $\eta_j$ is an elementary vector that is zero everywhere except for the dimension of the state $X_j^*(t)$ (\eqref{eq:delay_state}), that is $\eta_j^{\T}\mathbf{X}(t) = X_j^*(t)$. This representation is essential to the construction of our stochastic simulation schemes presented later. We note here that there are connections to the delay reaction setting presented by Anderson~\cite{Anderson2007} and Boguñá et al.,~\cite{Boguna2014}. The distinct properties of our approach by comparison to these previous works are discussed along with the derivation of our methods in subsequent sections.

\subsection{Simulation-based inference}
\label{subsec: 2.2 computational inference}

Suppose that we obtain time series data, $\data{obs} = [\mathbf{Y}(t_1),\mathbf{Y}(t_2),\ldots, \mathbf{Y}(t_n)]$, that is assumed to be noisy observations at $n$ discrete times, $t_1 < t_2, < \cdots < t_n$, from a realisation of  the stochastic process $\{\mathbf{X}(t)\}_{t > t_0}$ (\eqref{eqn:Kurtz} or \eqref{eq:kurtz_delay}). The aim is to infer the unknown rate parameters of the model $\theta = [\{k_j\}_{j \in \mathcal{R}}, \{k_j^*\}_{j \in \mathcal{R_D}}]$ via Bayes' theorem (\eqref{eq:bayes}). Typically inference will proceed according to the expectation, 
\begin{equation}
	\condexpect{f(\theta)}{\data{obs}} = \int_\Theta f(\theta)\condpdf{\theta}{\data{obs}}\,\text{d}\theta, \label{eq:postexp}
\end{equation}
for any $f(\theta)$ that is integrable with respect to the posterior measure with density $\condpdf{\theta}{\data{obs}}$. The form of \eqref{eq:postexp} is very general and includes estimation of means, variances, higher-order moments, probabilties, or probability densities of the posterior distribution through the specification of the function $f(\cdot)$. Assuming one can draw independent, identically distributed (i.i.d.) samples from the posterior , then we estimate using Monte Carlo (MC) integration,
\begin{equation*}
	\condexpect{f(\theta)}{\data{obs}} \approx \hat{f} = \frac{1}{N} \sum_{i=1}^N f(\theta^{(i)}),
\end{equation*}
where $\theta^{(1)},\theta^{(2)},\ldots,\theta^{(N)} \iid \condpdf{\cdot}{\data{obs}}$. Unfortunately, standard Bayesian sampling approaches that are based on MCMC or SMC are not viable since the likelihood
is intractable due to noisy partial observations and dependence on the master equation. This remains true even in the Markovian case where CME is substantially simpler solve (\eqref{eq:cme}). 

To handle this intractable likelihood setting, various simulation-based techniques can be used~\cite{Cranmer2020,Sisson2018}. Regardless of the specific method applied, the dominating computational cost is associated with stochastic simulations from the model. A clear example of this are ABC-based methods that approximate the posterior according to,
\begin{align*}
	\condpdf{\theta}{\data{obs}} \approx \pi_{\text{ABC}} \cond{\theta}{\rho\left(\data{obs},\data{sim} \right)< \varepsilon} \propto \condprob{\rho\left(\data{obs},\data{sim} \right)< \varepsilon}{\theta}\pdf{\theta},
\end{align*}
where 
\begin{equation*}
	\condprob{\rho\left(\data{obs},\data{sim} \right)< \varepsilon}{\theta} = \int \indd{(0,\varepsilon]}{\rho\left(\data{obs},\data{sim} \right)} \condpdf{\data{sim}}{\theta}\, \text{d}\data{sim}
\end{equation*}
with $\data{sim} \sim \condpdf{\cdot}{\theta}$ being simulated data through stochastic simulation, $\rho\left(\data{obs},\data{sim} \right)$ is a discrepancy metric, and $\varepsilon > 0$ is the acceptance threshold~\cite{Sisson2018,Sunnaaker2013,Warne2019}. 

Inference proceeds via expectations with respect to this approximation,
\begin{equation*}
	\condexpect{f(\theta_\varepsilon)}{\data{obs}} = \int_\Theta f(\theta_\varepsilon)\pi_{\text{ABC}} \cond{\theta_\varepsilon}{\rho\left(\data{obs},\data{sim} \right)< \varepsilon}\,\text{d}\theta_\varepsilon,
\end{equation*}
where $\theta_{\varepsilon} \sim \pi_{ABC} \cond{\theta}{\rho\left(\data{obs},\data{sim} \right)< \varepsilon}$.
In an accept-reject scheme \cite{Prescott2020} (Algorithm~\ref{alg: ABC rej}), i.i.d. samples from the joint prior-data distribution, $\pi(\data{sim},\theta) = \condpdf{\data{sim}}{\theta} \pi(\theta)$, are used to  enable the MC estimator,
\begin{equation}
	\condexpect{f(\theta_\varepsilon)}{\data{obs}} \approx \hat{f}_{\varepsilon} = \frac{\sum_{i=1}^N w(\theta^{(i)})f(\theta^{(i)})}{\sum_{i=1}^N w(\theta^{(i)})}, \label{eq:abc_mc}
\end{equation} 
where $\theta^{(1)},\theta^{(2)},\ldots,\theta^{(N)} \iid \pi(\cdot)$, and $w(\theta) = \indd{[0,\varepsilon)}{\rho\left(\data{obs},\data{sim} \right)}$. Here $\indd{A}{x}$ is the indicator function, $\indd{A}{x} = 1$ if $x \in A$ and $\indd{A}{x} = 0$ otherwise.  Due to the acceptance threshold $\varepsilon > 0$ and discrepancy based approximation, any expectations taken with respect to the ABC posterior are not exact as $\condexpect{f(\theta_\varepsilon)}{\data{obs}} \neq \condexpect{f(\theta)}{\data{obs}}$ in general. As a result, we ideally take $\varepsilon \to 0$, however, this leads to extremely low acceptance probabilities of $\mathcal{O}(\varepsilon^{d})$ where $d \geq 1$ is the dimensionality of the data, $\data{obs}$~\cite{Barber2015}. This curse of dimensionality renders Algorithm~\ref{alg: ABC rej} inefficient when implemented directly, however, extensions based on MCMC~\cite{Marjoram2003}, SMC~\cite{Drovandi2010,Sisson2007}, MLMC~\cite{Jasra2019,Warne2018} and multifideility schemes~\cite{Prescott2020} are widely applied in practice. 


\begin{algorithm}[h!]
	\caption{ABC accept-reject scheme (See Prescott \& Baker \cite{Prescott2020})}
	\begin{algorithmic}[l]
		\Require A stochastic model $\condpdf{\cdot}{\theta}$; a prior distribution $\pi(\theta)$; a discrepancy metric $\rho(\data{obs},\cdot)$; a discrepancy threshold $\varepsilon > 0$; observed data $y_{\mathrm{obs}}$; and sample size $N$.
		\For{$i =1,\dots, N$}
		\State Sample prior $\theta^{(i)} \sim \pi(\cdot)$;
		\State Generate simulated data $\data{sim} \sim \condpdf{\cdot}{\theta^{(i)}}$ ;
		\State $w^{(i)} \leftarrow \indd{[0,\varepsilon)}{\rho(\data{obs},\data{sim})}$;
		\EndFor
		\State \Return $\left(\theta^{(1)}, w^{(1)}\right), \left(\theta^{(2)}, w^{(2)}\right),\ldots,\left(\theta^{(N)}, w^{(N)}\right)$.
	\end{algorithmic}
	\label{alg: ABC rej}
\end{algorithm}

\subsubsection{Multifidelity approximate Bayesian computation}
\label{subsec: 2.2.2 MF-ABC}

Prescott and Baker~\cite{Prescott2020} consider an approach to accelerate the computation of $\hat{f}_{\varepsilon}$ (\eqref{eq:abc_mc}) that is based upon the probabilistic telescoping summation idea of Rhee and Glynn~\cite{Rhee2015}. Here, we denote $\approxcondpdf{\cdot}{\theta}$ as an approximate simulator that is computationally inexpensive to generate a sample $\approxdata{sim}$ from $\approxcondpdf{\cdot}{\theta}$. We refer to this approximation as a \textit{low-fidelity} simulator, and the exact simulator, $\condpdf{\cdot}{\theta}$, as a \textit{high-fidelity} simulator that is computationally expensive to generate a sample $\data{sim}$ from $\condpdf{\cdot}{\theta}$. Direct substitution of the low-fidelity simulator into Algorithm~\ref{alg: ABC rej} will be unreliable due to propagation of simulation error into posterior estimation error~\cite{Prescott2024}. This problem motivates the multifidelity approach that exploits the efficiency of the low-fidelity simulator while maintaining the accuracy of the high-fidelity simulator.  

In the context of ABC, a so called multifidelity weighting function is introduced,
\begin{align}
	w_{\mathrm{mf}}(\theta) = \indd{[0,\varepsilon)}{\rho(\data{obs},\approxdata{sim})} + \frac{m}{\mu(\approxdata{sim})}\left[\indd{[0,\varepsilon)}{\rho(\data{obs},\data{sim})} - \indd{[0,\varepsilon)}{\rho(\data{obs},\approxdata{sim})}\right], \label{eq:mf_w}
\end{align}
where $m$ is the outcome of a Bernoulli random variable, $M$, such that $\condprob{M = 1}{\approxdata{sim}} = \mu(\approxdata{sim})$. The function, $\mu(\approxdata{sim})$, represents the probability of computing a correction of $\pm 1/\mu(\data{sim})$ when\\ $\indd{[0,\varepsilon)}{\rho(\data{obs},\data{sim})}  \neq \indd{[0,\varepsilon)}{\rho(\data{obs},\approxdata{sim})}$. Note that if $m = 0$, then $w_{\mathrm{mf}}(\theta) = \indd{[0,\varepsilon)}{\rho(\data{obs},\approxdata{sim})}$ and the high-fidelity simulator is not needed. Using this weighting scheme, we can construct the multifidelity estimator,
\begin{equation}
	\hat{f}_{\mathrm{mf},\varepsilon} = \frac{\sum_{i=1}^N w_{\mathrm{mf}}(\theta^{(i)})f(\theta^{(i)})}{\sum_{i=1}^N w_{\mathrm{mf}}(\theta^{(i)})}, \label{eq:mf_is}
\end{equation}
where $\theta^{(1)},\theta^{(2)},\ldots, \theta^{(N)} \iid \pi(\cdot)$. It can be shown that $\hat{f}_{\mathrm{mf},\varepsilon}$ is a consistent estimator of $\condexpect{f(\theta_\varepsilon)}{\data{obs}}$~\cite{Prescott2020,Prescott2024}. Furthermore, given a fixed computational budget, one can choose a mean function, $\mu(\approxdata{sim})$, such that $\var{\hat{f}_{\text{mf},\varepsilon}} < \var{\hat{f}_\varepsilon}$, provided certain conditions hold~\cite{Prescott2024} that relate to the accuracy of the low-fidelity simulator relative to the computational gains. Broadly speaking, \eqref{eq:mf_is} utilises the high-fidelity model to correct bias introduced by the low-fidelity model through the weighting scheme (\eqref{eq:mf_w}).

For simplicity we will restrict our description to the special case where the decision to  simulate $\data{sim} \sim \condpdf{\cdot}{\theta}$ is informed only by the acceptance decision of a given low-fidelity simulation $\approxdata{sim} \sim \approxcondpdf{\cdot}{\theta}$. That is,
\begin{align}
	\mu(\approxdata{sim}) = \mu_a\indd{[0,\varepsilon)}{\rho(\data{obs},\approxdata{sim})} + \mu_r\indd{[\varepsilon, \infty)}{\rho(\data{obs},\approxdata{sim})}, \label{eq:mf_mu}
\end{align}
where $\mu_a \in [0,1] $ and $\mu_r \in [0,1]$ are the continuation probabilities when $\approxdata{sim}$ is accepted or, respectively, rejected. Assuming $\mu_a$ and $\mu_r$ are given, \eqref{eq:mf_w}--(\ref{eq:mf_mu}) can be used to implement a multifidelity ABC (MF-ABC) accept-reject scheme as shown in Algorithm~\ref{alg: MF-ABC}. To achieve optimal performance, Prescott and Baker~\cite{Prescott2020} perform a training step for a set of $N_w$ warm-up simulation pairs $(\approxdata{sim}^{(1)},\data{sim}^{(1)}), (\approxdata{sim}^{(1)},\data{sim}^{(1)}), \ldots, (\approxdata{sim}^{(N_w)},\data{sim}^{(N_w)})$. This is extended to an adaptive tuning of $\mu_a$ and $\mu_r$ in Warne et al.,~\cite{Warne2022}, then in Prescott et al.,~\cite{Prescott2024} a $\theta$ dependent version of the continuation probability function is developed. 

In general, positively correlated simulation pairs are essential to obtain substantial performance gains (Appendix~\ref{app:contprob}). 
In the Markovian setting, MF-ABC inference and various extensions can be directly implemented using the exact stochastic simulation schemes of Gillespie~\cite{Gillespie1977} and Anderson~\cite{Anderson2007}, the approximate stochastic simulation scheme of Gillespie~\cite{Gillespie2001} and the coupling scheme of Anderson and Higham~\cite{Anderson2012}. However, to-date the literature has primarily focused on exact stochastic simulation schemes for a non-Markovian reaction networks~\cite{Anderson2007,Boguna2014,Cai2007,Fu2022,Voliotis2016}. Two notable exceptions are the neural network aided approximation scheme of Jiang et al.~\cite{Jiang2021} and the recent scalable approach of P\'{e}lissier et al.,~\cite{Pelissier2026}. While both approaches are highly effective, neither approach leads to an exact coupling scheme that is required to implement multifidelity methods.
This leads to our main contribution, that is, exact and approximate stochastic simulation algorithms with appropriate coupling schemes for non-Markovian chemical reaction networks. These contributions are presented in the follow subsections. 

\begin{algorithm}[H]
	\caption{Multifidelity ABC accept-reject scheme}
	\begin{algorithmic}[l]
		\Require A low-fidelity simulator $\approxcondpdf{\cdot}{\theta}$; a high-fidelity simulator $\condpdf{\cdot}{\theta}$; a prior distribution $\pi(\theta)$; a discrepancy function $\rho(\data{obs},\cdot)$; a discrepancy threshold $\varepsilon$; observed data $\data{obs}$; a sample size $N$; and continuation probability function, $\mu(\cdot)$.
		\For{$i =1,\dots, N$}
		\State Generate a proposal $\theta^{(i)} \sim \pi(\theta)$;
		\State Generate low-fidelity simulation, $\approxdata{sim} \sim \approxcondpdf{\cdot}{\theta^{(i)}}$;
		\State Set $\tilde{w}^{(i)} \leftarrow \indd{[0,\varepsilon)}{\rho(\data{obs},\approxdata{sim})}$;
		\State Sample $M \sim \text{Bernoulli}(\mu(\approxdata{sim}))$; 
		\If{$M = 1$}
		\State Generate high-fidelity simulation, $\data{sim} \sim\condpdf{\cdot}{\theta}$;
		\State Set $w^{(i)} \leftarrow \indd{[0,\varepsilon)}{\rho(\data{obs},\data{sim})}$;
		\State Set $w_{\text{mf}}^{(i)} \leftarrow \tilde{w}^{(i)} + \left(w^{(i)} - \tilde{w}^{(i)}\right)/\mu(\approxdata{sim}) $;
		\Else
		\State $w_{\text{mf}}^{(i)} \leftarrow \tilde{w}^{(i)}$;
		\EndIf
		\EndFor
		\State \Return $\left(\theta^{(1)}, w_{\text{mf}}^{(1)}\right), \left(\theta^{(2)}, w_{\text{mf}}^{(2)}\right),\ldots,\left(\theta^{(N)}, w_{\text{mf}}^{(N)}\right)$.
	\end{algorithmic}
	\label{alg: MF-ABC}
\end{algorithm}

\subsection{Stochastic simulation algorithms for non-Markovian processes}
\label{subsec: 2.3 Stochastic Simulation}
Stochastic simulation schemes for Markovian reaction networks have been well studied~\cite{Gillespie2013,Higham2008,Schnoerr2017,Warne2019} with many exact and approximate stochastic simulation algorithms available. In addition, exact schemes for handling certain time-dependent propensity functions and certain types of delays are also available. For example, Anderson's modified NRM~\cite{Anderson2007} handles time dependent propensities via numerical integration, and the Extrande method of Voliotis et al.,~\cite{Voliotis2016} provides an efficient alternative based on thinning techniques. Anderson also considers delayed reactions with either fixed completion times, or random completion times that can depend on time~\cite{Anderson2007}. However, Boguñá et al.,~\cite{Boguna2014} note that the Anderson approach cannot directly be extended to the setting where delays are random and depend on both time and system state. To deal with this Boguñá et al.,~\cite{Boguna2014} extend the Gillespie method though numerical inversion of the next reaction time in this general setting. By constrast, there have been relatively few works to date on approximate stochastic simulation schemes of non-Markovian systems~\cite{Jiang2021,Pelissier2026}. For the existing approximations, there is no natural exact coupling scheme that arises. As result, application of MLMC and multifidelity schemes is not achievable without a new approximation scheme.

In this section, we present our main contributions: i) An exact stochastic simulation algorithms for non-Markovian reaction networks of the general form with state and time dependent delays. This is achieved through a novel generalisation to the Anderson modified NRM~\cite{Anderson2007} for fixed delays; ii) an approximate stochastic simulation algorithm that is based on a $\tau$-leaping discretisation of our non-Markovian modified NRM; then iii) we extend the exact coupling scheme of similar form to Anderson and Higham~\cite{Anderson2012}, Lester et al.~\cite{Lester2016}, and Prescott et al.~\cite{Prescott2024}, to generate positively correlated exact and approximate simulation pairs in the non-Markovian case. 

\subsubsection{Exact stochastic simulation algorithms}
\label{subsec: 2.3.1 ESSA}

The NRM~\cite{Cao2004,Gibson2000} is an alternative exact stochastic simulation algorithm to the Gillespie direct method~\cite{Gillespie1977}. Both methods generate sample paths that exactly follow the probability law of a discrete-state continuous-time Markov process given in \eqref{eqn:Kurtz}. The Gillespie direct method operates through sampling the global waiting time to the next reaction and then sampling the next event conditional on this waiting time. In contrast, the NRM is derived directly from the Kurtz random time-change representation and proceeds through maintaining independent unit rate Poisson processes for each reaction channel and determining the next even via the minimum of the waiting times. This typically offers a computational advantage as only one random variable is required per event. For non-Markovian processes, such as those presented in Section \ref{subsec: 2.1.2 slow reactions}, the equivalent approach to the Gillespie direct method, that is the non-Markovian Gillespie Algorithm (nM-GA) of Boguñá et al.,~\cite{Boguna2014}, requires additional approximations in practice since the cumulative hazard function for the process rarely admits a closed-form quantile function required to sample the global waiting time. 

We extend the NRM to the non-Markovian setting by considering a race of Poisson processes with time and history dependent cumulative hazard functions. That is the next reaction time of each reaction channel is computed independently assuming the state vector $\mathbf{X}(t)$ remains unchanged. To proceed, we first define the cumulative hazard function for instantaneous reactions, $j \in \mathcal{R}_I$, and delay initiations, $j \in \mathcal{R}_D$, as
\begin{equation}
		\Lambda_j(\tau; \mathbf{X}(t), t) = \int_0^\tau  \lambda_j(\mathbf{X}(t+s),t+s)\,\text{d}s. 
	\label{eq:chfi}
\end{equation}
Similarly, we define the cumulative hazard function for delay completions, $j \in \mathcal{R}_{D}$,
\begin{equation}
		\Lambda_j^*(\tau ; \mathbf{X}(t),t, a) = \int_0^{\tau} \lambda^*_j(\mathbf{X}(t+s),t - a +s)\,\text{d}s,\label{eq:chfd}
\end{equation}
where $a > 0$ is some time-shift based on the history of the reaction network.
Given the definitions in \eqref{eq:chfi}--(\ref{eq:chfd}), we observe that the next event time for a instantaneous reaction $j \in \mathcal{R}_I$ or the initiation of a delayed reaction $j \in \mathcal{R}_D$, is given by solving
\begin{equation}
	\tau_{j,u} = 	\Lambda_j(\tau_j; \mathbf{X}(t), t),
	\label{eq:nri}
\end{equation}
for $\tau_j$, where $\tau_{j,u}$ is the next event time for the unit rate Poisson process $Y_j(\cdot)$, which is an exponentially distributed random variable with unit rate, $\tau_{j,u} \sim \text{Exp}(1)$ as determined under the Kurtz random time-change formulation (\eqref{eq:kurtz_delay}). Similarly, the completion of the $k$th delayed reaction event, $j \in \mathcal{R}_D$ with initiation times $t_{j,0}^{*,k}$, the completion time is obtained by solving 
\begin{equation}
	\tau^k_{j,u} = \Lambda_j^*(\tau^k_j ; \mathbf{X}(t),t, t_{j,0}^{*,k}) 
	\label{eq:nrd}
\end{equation}
for $\tau_j^k$, where $\tau^k_{j,u}$ is the next event time for the unit rate Poisson process $Y^k_j(\cdot)$. 

Note that for $t - t_{j,0}^{*,k} + s < 0$ then we define  $\lambda^*_j(\mathbf{X}(t+s),t - t_{j,0}^{*,k}+s) = 0$, that is, a completion event cannot occur before it is initiated. Given the set of next event times for each channel independently, we can obtain the overall next event time for the system via $\tau_{\text{min}} = \min{(\{\tau_j\}_{j \in \mathcal{R}} \cup \{\tau_j^k\}_{j \in \mathcal{R}_D, k \in \mathcal{D}_j})}$ where $\mathcal{D}_j$ is the set of delayed reactions yet to be completed, with $|\mathcal{D}_j| = X_j^*(t)$. The specific value of $j$ and $k$ that is associate with $\tau_{\mathrm{min}}$ will be the next event that occurs.

Equations (\ref{eq:nri}) and (\ref{eq:nrd}) provide the \textit{first reaction} time only, given the time and state dependence. Thus for all reactions channels other than the one that fired we need to solve for the remaining time left, $r_{j,u} > 0$, in the time interval of the unit rate Poisson process following the state update. Focusing on instantaneous and delay initiation channels, at time $t > t_0$ we denote the internal time, that is the time since the last reaction, as $t_j$. We have at time $t > t_0$, for all reactions, 
\begin{align}
	\tau_{j,u} = r_{j,u} + \Lambda_j(t_j + \tau_{\text{min}}; \mathbf{X}(t), t) - \Lambda_j(t_j; \mathbf{X}(t), t).\label{eq:rtd} 
\end{align}
Note that the firing reaction will, according to \eqref{eq:rtd}, result in $r_{j,u} = 0$. In this case, a new remaining time interval for the unit rate Poisson process is generated, that is, $r_{j,u} = \tau_{j,u} \sim \mathrm{Exp}(1)$. 

\eqref{eq:rtd} leads to an update rule for $r_{j,u}$ following any instantaneous reaction event,
\begin{equation}
	r_{j,u} \leftarrow r_{j,u} - \left[\Lambda_j(t_j + \tau_{\text{min}}; \mathbf{X}(t), t) - \Lambda_j(t_j; \mathbf{X}(t), t)\right]. \label{eq:rit}
\end{equation}
For the completion of the $k$th delay event, the equivalent update would similarly be,  
\begin{equation}
	r^k_{j,u} \leftarrow r^k_{j,u} - \left[\Lambda^*_j(t_j^k + \tau_{\text{min}}; \mathbf{X}(t), t, t_{j,0}^{*,k}) - \Lambda^*_j(t_j^k; \mathbf{X}(t), t, t_{j,0}^{*,k})\right]. \label{eq:ritd}
\end{equation}
Given remaining intervals and internal times enables the update of next reaction times. This is given for the instantaneous and initiation reactions by solving,
\begin{equation} 
	\Lambda_j(\tau_j + t_j; \mathbf{X}(t), t) = r_{j,u} + \Lambda_j(t_j; \mathbf{X}(t), t), \label{eq:nrt}
\end{equation}
for $\tau_j$, and for the delay completions by solving 
\begin{equation}
	\Lambda^{*}_j(\tau^k_j + t^k_j ; \mathbf{X}(t), t,t_{j,0}^{*,k}) = r^k_{j,u} + \Lambda^*_j(t^k_j; \mathbf{X}(t), t,t_{j,0}^{*,k}). \label{eq:nrtd}
\end{equation}
for $\tau^k_j$. 

Equations (\ref{eq:nri})--(\ref{eq:nrtd}) provide a mechanism for updating next reaction times and remaining interval lengths in the general setting of time and state dependent propensities and including time delays. This enables direct simulation of the inhomogeneous Poisson processes in \eqref{eq:kurtz_delay}. Unlike the Markovian case, the remaining interval lengths \eqref{eq:rit}--(\ref{eq:ritd}) must be updated for every reaction channel following every reaction event. In settings with large numbers of concurrent incomplete delay channels, a global bounding scheme on the instantaneous rates, following P\'{e}lissier et al.,~\cite{Pelissier2026}, could be implemented at the cost of some accuracy.

To track ongoing delays within reaction channels $j\in \mathcal{R}_D$ we maintain a set $\mathcal{D}_j \subset \mathbb{N}$ that contains the $k$ indices for initiated delays that are yet to be completed. For any $k \in \mathbb{N}$, the only completion processes that need to be simulated are the $Y_j^k(\cdot)$ such that $k \in \mathcal{D}_j$. For such as process, we only need to simulate to the first event time of $Y_j^k(\cdot)$ as we take  $\min{(1, Y_j^k(\cdot))}$, therefore after the completion event and $Y_j^k(\cdot) = 1$, we remove $k$ from $\mathcal{D}_j$. Similarly, at the completion of a delayed reaction cannot occur before the initiation event, that is, $\lambda_j^*(\mathbf{X}(t), t- t_{j,0}^{*,k}) = 0$ when $t < t_{j,0}^{*,k}$.  Thus, when the $k$th initiation event occurs, the index $k$ is added to $\mathcal{D}_j$. As a result of this, the number of elements in the set $\mathcal{D}_j$ is always the same as the copy number for the ongoing reactions $X_j^*(t)$ (\eqref{eq:delay_state}) at any time $t >t_0$.  

Combining the direct implementation of the Kurtz random time-change representation with the dynamic tracking of progressing delayed reactions via the sets $\mathcal{D}_j$ for $j \in \mathcal{R}_D$ leads to a non-Markovian next reaction method (nM-NRM). The method, presented in Algorithm~\ref{alg: nm-nrm}, is an extension of Anderson's NRM that dealt with fixed delays~\cite{Anderson2007} to enable random delays that depend on both time and system state. This means we achieve the same generality as the non-Markovian Gillespie method of Boguñá et al.,~\cite{Boguna2014}. However, our approach leads to convenient approximation and coupling schemes that are essential for efficient sampling based on MLMC or multifidelity methods such as the multifidelity ABC method~(Algorithm~\ref{alg: MF-ABC}). In particular, the random time-change representation allows all necessary interval lengths $r_{j,u}$ and $r_{j,u}^k$ for the unit rate Poisson processes to be generated prior to the start of a simulation. This provides direct analogues of the $\tau$-leaping schemes~\cite{Gillespie2001} and coupling based on common Poisson clocks~\cite{Anderson2012,Prescott2020}. 

\begin{algorithm}[H]
	\caption{Non-Markovian next reaction method (nM-NRM)}
	\begin{algorithmic}[l]
		\Require A non-Markovian reaction network with $\mathcal{N}$ species and $\mathcal{M}$ reaction channels, of which $\mathcal{D}$ are delay reactions (\eqref{eq:delay_bcrn}); an initial time $t_0\geq0$; a final time $T > t_0$; an initial system state $\mathbf{x}_0=\mathbf{X}(t_0)$;
		\State Initialise state $\mathbf{X} \leftarrow \mathbf{x}_0$, time $t \leftarrow t_0$, internal times $t_j \leftarrow 0$ for $j \in \mathcal{R}$, next reaction intervals, $r_{j,u} \sim \text{Exp}(1)$ for $j \in \mathcal{R}$, and delay sets $\mathcal{D}_j \leftarrow \emptyset$, $j \in \mathcal{R}_D$;
		\While{$t < T$}
		\State Update reaction times $\tau_{j}$ for $j \in \mathcal{R}_I$, and $\tau^k_{j}$ for $k \in \mathcal{D}_j$, $j \in \mathcal{R}_D$ ((\ref{eq:nrt})--(\ref{eq:nrtd}));
		\State Find $\tau_{\mu_1} \leftarrow \min_{j \in \mathcal{R}}{(\tau_j)}$, and $\tau_{\mu_2}^k \leftarrow \min_{j \in \mathcal{R}_D, k \in \mathcal{D}_j}{(\tau_j^k)}$;
		\State Set next event time $\tau_{\text{min}} \leftarrow \min{(\tau_{\mu_1},\tau_{\mu_2}^k)}$
		\State Update intervals $r_{j,u}$ for $j \in \mathcal{R}$, and $r_{j,u}^k$ for $k \in \mathcal{D}_j$, $j \in \mathcal{R}_D$ ((\ref{eq:rit})--(\ref{eq:ritd}));
		\State Update internal times $t_{j} \leftarrow t_{j} + \tau_{\text{min}}$ for $j\in \mathcal{R}$, and $t_{j}^k \leftarrow t_{j}^k + \tau_{\text{min}}$ for $k \in \mathcal{D}_j$ $j \in \mathcal{R}_D$;
		\If{$\tau_{\text{min}} = \tau_{\mu_1}$}
		\State Set $r_{\mu_1,u} \sim \text{Exp}(1)$;
		\If{$\mu_1 \in \mathcal{R}_I$}
		\State Set $\nu \leftarrow \nu_{*,\mu_1}$;
		\Else
		\State Set $\nu \leftarrow \eta_{\mu_1} -\nu_{*,\mu_1}^{-}$;
		\State Set $k' \leftarrow X^*_{\mu_1}+ 1$;
		\State Set $t_{\mu_1,0}^{*,k'} \leftarrow t + \tau_{\mu_1}$, $t_{\mu_1}^{k'} \leftarrow 0$, and $r_{\mu_1,u}^{k'} \sim \text{Exp}(1)$ 
		\State Initiate delayed reaction $\mathcal{D}_{\mu_1} \leftarrow \mathcal{D}_{\mu_1} \cup \{k'\}$;
		\EndIf
		\Else
		\State Set $\nu \leftarrow \nu_{*,\mu_2}^+ -\eta_{\mu_2}$;
		\State Complete delayed reaction $\mathcal{D}_{\mu_2} \leftarrow \mathcal{D}_{\mu_2} \backslash \{k\}$;
		
		\EndIf
		\State Update the system state $\mathbf{X} \leftarrow \mathbf{X} + \nu$, and time $t \leftarrow t + \tau_{\text{min}}$;
		\EndWhile
	\end{algorithmic}
	\label{alg: nm-nrm}	
\end{algorithm}

\subsubsection{Approximate stochastic simulation algorithms}
\label{sec: 2.3.2 nm-tl}

The $\tau$-leaping method introduced by Gillespie~\cite{Gillespie2001} is an approximate stochastic simulation algorithm that efficiently simulates reaction networks by discretising time and allowing for multiple reaction events per time step. The $\tau$-leaping method is widely used in the simulation of reaction networks where exact methods like the Gillespie algorithm are prohibitively expensive~\cite{Higham2008,Warne2019}. While many $\tau$-leaping variants exist, to our knowledge, none are suitable for non-Markovian reaction networks where reaction propensities may depend on internal clocks or history. Leveraging the general framework for non-Markovian reaction network with delays, given by~\eqref{eq:kurtz_delay}, we develop here an extension to the $\tau$-leaping algorithm to approximate sample paths.

The nM-NRM (Algorithm~\ref{alg: nm-nrm}) proceeds by explicitly simulating every reaction event that occurs. As a result, the computational cost depends heavily on the global average reaction rate. The $\tau$-leaping method mitigates this by invoking a simplifying assumption that the system state changes very little over the interval $[t, t+\tau)$, that is $\mathbf{X}(t+ s) \approx \mathbf{X}(s)$ for $s \in [0,\tau)$. 

We denote $\mathbf{Z}(t) = [Z_1(t),Z_2(t),\ldots,Z_\mathcal{N}(t),Z_{j_1}^*(t),Z_{j_2}^*(t),\ldots, Z_{j_{\mathcal{D}}}^*(t)]$, as the approximation to $\mathbf{X}(t)$ that arises by assuming $\mathbf{Z}(t+s) = \mathbf{Z}(t)$ for $s \in [0,\tau)$. This leads to an approximation of the cumulative hazard functions,
\begin{equation}
	\begin{split}
		\Lambda_j(\tau; \mathbf{X}(t), t) &\approx \tilde{\Lambda}_j(\tau; \mathbf{Z}(t), t) = \int_0^\tau  \lambda_j(\mathbf{Z}(t),t+s)\,\text{d}s, \\
		\Lambda_j^*(\tau ; \mathbf{X}(t),t, a) &\approx \tilde{\Lambda}_j^*(\tau ; \mathbf{Z}(t),t, a) = \int_0^{\tau} \lambda^*_j(\mathbf{Z}(t),t - a +s)\,\text{d}s.
	\end{split}\label{eq:chfa}
\end{equation}
By substitution of \eqref{eq:chfa} into \eqref{eq:kurtz_delay} we obtain
\begin{equation}
	\mathbf{Z}(t+\tau) = \mathbf{Z}(t) + \sum_{j\in \mathcal{R}_I}P_j\nu_{*,j} +\sum_{j\in \mathcal{R}_D}P_j(\eta_j-\nu^-_{*,j}) + \sum_{j\in \mathcal{R}_D}\sum_{\kappa=1}^\infty \min{\left\{Z_j^{*,\kappa}(t),P^\kappa_j\right\}}(\nu^+_{*,j}-\eta_j),   
	\label{eq:tau_leap_update}
\end{equation} 
where $P_j \sim \text{Poisson}\left(\tilde{\Lambda}_j(\tau;\mathbf{Z}(t),t)\right)$ and $P^\kappa_j \sim \text{Poisson}\left(\tilde{\Lambda}^*_j(\tau;\mathbf{Z}(t),t,t_{j,0}^{*,\kappa})\right)$ are independent Poisson random variates. Note that in most cases the integrals in \eqref{eq:chfa} are analytically tractable. Unlike \eqref{eq:kurtz_delay} where every delayed reaction initiation has a unique initiation time $t_{j,0}^{*,k}$, the initiation process described by \eqref{eq:tau_leap_update}, for $j \in \mathcal{R}_D$, will initiate delays in groups of size $P_j > 0$. Within this group, each delay will have the same initiation time of $t_{j,0}^{*,\kappa} = t + \tau$. Note, we adopt the notation $t_{j,0}^{*,\kappa}$ for the initiation time of the $\kappa$th group in the time-discretised approximation, this is not to be confused with the notation of $t_{j,0}^{*,k}$ for the exact initiation times in the continuous-time process in \eqref{eq:kurtz_delay}. The grouping occurs due to the time discretisation. That is, over time interval $[t, t+ \tau)$ exactly $P_j \sim \text{Poisson}\left(\tilde{\Lambda}_j(\tau;\mathbf{Z}(t),t)\right)$ initiation events occur and enter the system in bulk at time $t_{j,0}^{*,\kappa} = t + \tau$ which is the next time-step. As a result, all of these $P_j$ delayed chemical species are identical and can be treated with the same reaction channel. That is at the $\kappa$th initiation group, we set $Z_j^{*,\kappa} = P_j$ for $j \in \mathcal{R}_D$ and include $\kappa$ in the delay group set $\mathcal{D}_j$. This new reaction channel remains in the system until it has fired a total of $Z_j^{*,\kappa}$ times, after which $\kappa$ is removed from the delay set $\mathcal{D}_j$. This leads to the non-Markovian $\tau$-leaping method (nM-TLM) in Algorithm~\ref{alg: nm-tl} .

\begin{algorithm}[H]
	\caption{Non-Markovian $\tau$-leaping method (nM-TLM)}
	\begin{algorithmic}[l]
		\Require A non-Markovian reaction network with $\mathcal{N}$ species and $\mathcal{M}$ reaction channels, of which $\mathcal{D}$ are delay reactions (\eqref{eq:delay_bcrn}); an initial time $t_0\leq0$; a final time $T > t_0$; an initial approximate system state $\mathbf{z}_0=\mathbf{Z}(t_0)$; a time discretisation $\tau$;
		\State Initialise state $\mathbf{Z} \leftarrow \mathbf{z}_0$, time $t \leftarrow t_0$, internal times $t_j \leftarrow 0$ for $j \in \mathcal{R}$, and delay sets $\mathcal{D}_j \leftarrow \emptyset$, $j \in \mathcal{R}_D$;
		\While{$t < T$}
		\State Evaluate $\tilde{\Lambda}_j(\tau;\mathbf{Z}(t),t)$ for $j \in \mathcal{R}$, and $\tilde{\Lambda}^*_j(\tau;\mathbf{Z}(t),t,t_{j,0}^{*,\kappa})$ for $\kappa \in \mathcal{D}_j$ and $j \in \mathcal{R}_D$;
		\State Sample $P_{j} \sim \mathrm{Poisson}\left(\tilde{\Lambda}_j(\tau;\mathbf{Z}(t),t)\right)$ for $j \in \mathcal{R}$;
		\State Sample  $P^\kappa_{j} \sim \mathrm{Poisson}\left(\tilde{\Lambda}^*_j(\tau;\mathbf{Z}(t),t_{j,0}^{*,\kappa})\right)$ for $\kappa \in \mathcal{D}_j$ and $j \in \mathcal{R}_D$
		\State Update system state using \eqref{eq:tau_leap_update};
		\State Update delayed species $Z_j^{*,\kappa} \leftarrow \max{(0,Z_j^{*,\kappa} - P_j^\kappa)}$ for $\kappa \in \mathcal{D}_j$ and $j \in \mathcal{R}_D$;
		\State Remove completed delay channels $\mathcal{D}_j \leftarrow \mathcal{D}_j \backslash \{\kappa : Z_j^{*,\kappa} = 0\}$ for  $j \in \mathcal{R}_D$;
		\For{$j \in \mathcal{R}_D$ such that $P_j > 0$}
		\State Set $\kappa' \leftarrow Z_j^* + 1$
		\State Set $t_{j,0}^{*,\kappa'} \leftarrow t +\tau$ and $Z_j^{*,\kappa'} \leftarrow P_j$ for  $j \in \mathcal{R}_D$;
		\State Update $\mathcal{D}_j \leftarrow \mathcal{D}_j \cup \{\kappa'\}$
		\EndFor
		\State Update time $t \leftarrow t +\tau$;
		\EndWhile
	\end{algorithmic}
	\label{alg: nm-tl}
\end{algorithm}

While we do not analytically consider the rate of convergence, it is straightforward to show using the same reasoning as Gillespie~\cite{Gillespie2001} that  \eqref{eq:tau_leap_update} converges to \eqref{eq:kurtz_delay} as $\tau \to 0$. We report numerical results for both weak and strong convergence rates in the \nameref{sec: 3. Results} Section.


\subsubsection{Unbiased multilevel Monte Carlo}
Given the exact and approximate stochastic simulation schemes (Algorithms~\ref{alg: nm-nrm} and \ref{alg: nm-tl}), we can develop a coupling scheme that generates an exact simulation that is driven by the same inhomogeneous Poisson processes underlying a given approximate simulation (See \ref{app:coupled-nM-NRM} for details). Using this coupling scheme we can implement an unbiased MLMC estimator following the approaches of Anderson and Higham~\cite{Anderson2012} and Lester et al.,~\cite{Lester2016}. In this setting, we aim to estimate the expectation,
\begin{equation*}
	\expect{f(\mathbf{X}(T))} = \sum_{\mathbf{x}_T \in \mathbb{N}^\mathcal{N}} f(\mathbf{x}_T) \condprob{\mathbf{X}(T) =  \mathbf{x}_T}{\mathbf{X}(t_0) = \mathbf{x}_0},
\end{equation*}  
where $f(\cdot)$ is some function of the system state, and $\condprob{\mathbf{X}(T) =  \mathbf{x}_T}{\mathbf{X}(t_0) = \mathbf{x}_0}$ is the solution to the forward problem of the non-Markovian network at time $t = T$, given an initial state $\mathbf{X}(t_0) = \mathbf{x}_0$. We expand the expectation using the telescoping summation,
\begin{equation*}
	\expect{f(\mathbf{X}(T))} = \expect{f(\mathbf{Z}(T))} + \expect{f(\mathbf{X}(T)) - f(\mathbf{Z}(T))},
\end{equation*}
where $\mathbf{Z}(T)$ is the approximate stochastic process as determind by \eqref{eq:tau_leap_update} for a given discretisation $\tau$. This leads to the MLMC estimator
\begin{equation}
	\hat{f}_{\mathbf{X}} = \frac{1}{N_0}\sum_{i=1}^{N_0} f(\mathbf{Z}^{(i)}(T)) + \frac{1}{N_1}\sum_{i=1}^{N_1}f(\mathbf{X}^{(i)}(T)) - f(\mathbf{Z}^{(i)}(T)). \label{eq:mlmc_est}
\end{equation}
Here, $\mathbf{Z}^{(1)}(T),\mathbf{Z}^{(2)}(T), \ldots, \mathbf{Z}^{(N_0)}(T)$ are i.i.d. samples from the nM-TLM (Algorithm~\ref{alg: nm-tl}) and \\$(\mathbf{X}^{(1)}(T),f(\mathbf{Z}^{(1)}(T)), (\mathbf{X}^{(2)}(T),f(\mathbf{Z}^{(2)}(T)), \ldots, (\mathbf{X}^{(N_1)}(T),f(\mathbf{Z}^{(N_1)}(T))$ i.i.d. coupled nM-NRM and nM-TLM pairs (Algorithm~\ref{alg: nm-tl} and Appendix~\ref{app:coupled-nM-NRM}). For a target mean square error (MSE), $h^2$, the optimal choice of sample sizes, $N_\ell$ for $\ell = 0, 1$,  is given by 
\begin{equation*}
N_\ell \propto \frac{1}{h^2}\sqrt{\frac{V_\ell}{C_\ell}}\left[\sqrt{V_0 C_0} + \sqrt{V_1 C_1}\right]	
\end{equation*}
where $V_0 = \var{f(\mathbf{Z}(T)}$ and $V_1 = \var{f(\mathbf{X}(T) - f(\mathbf{Z}(T)}$, and $C_0$ and $C_1$ are the expected computational cost (i.e., total simulation runtime) of a draw from $\mathbf{Z}(T)$ and $(\mathbf{X}(T),\mathbf{Z}(T))$, respetively. Importantly, since we can generate exact couplings between $\mathbf{X}(T)$ and $\mathbf{Z}(T)$, the resulting MLMC estimator (\eqref{eq:mlmc_est}) is unbiased regardless of the choice of discretisation $\tau$.   

\section{Results}
\label{sec: 3. Results}

In this section, we demonstrate the performance and applicability of the algorithms introduced in Section 2. We begin by specifying two representative examples of non-Markovian reaction networks that include both delays and history-dependent dynamics. Using these models, we demonstrate sample paths generated by our algorithms and evaluate the accuracy and efficiency of the nM-TLM approximation. We then apply our exact coupling scheme to demonstrate accelerated simulation and inference using unbiased MLMC simulation~\cite{Anderson2012,Giles2008} and multifidelity ABC for inference~\cite{Prescott2020}. Our results show substantial estimator efficiency improvements of an order of magnitude are feasible using our methods compared with direct MC sampling.

\subsection{Example non-Markovian reaction networks}
We consider two examples of non-Markovian biochemical reaction networks to demonstrate the efficacy of our methods. The first is is a genetic regulation network that includes auto-inhibition with oscillatory dynamics arising from a delay in messenger RNA transcription~\cite{Barrio2006,Jensen2003}. The second is a non-Markovian extension to the susceptible-infectious-susceptible (SIS) model from epidemiology, in which both the infection and recovery processes have Weibull waiting time distributions.  

\subsubsection{Example 1: Gene regulation with delayed auto-inhibition}
\label{subsec: 3.1 gene regulation with delayed autoinhibition}
 The model describes the transcription of a gene coding DNA sequence into an messenger RNA (mRNA) molecule and subsequent translation of mRNA to proteins that, in-turn, auto-inhibit the transcription process. The process of transcription and subsequent transport of mRNA from cell nucleus to cytoplasm is non-negligible~\cite{Brett2013,Jensen2003}, thus we define the transcription process as a delay reaction. We also consider the resources required for transcription to be finite~\cite{Boguna2014}, therefore the processing time slows as the total number of mRNA in the increases, requiring a state-dependent completion function. Such delay models are of interest in the study of gene regulation as they are capable of oscillatory behaviour arising from auto-inhibitory mechanism, such as is observed in the Hes1 system associated with cell differentiation~\cite{Hirata2002,Jensen2003}. 


The delayed auto-inhibition model is a non-Markovian reaction network consisting of one delayed reaction, and three instantaneous reactions. The network is given by,
\begin{equation}
	\begin{split}
		\underbrace{\emptyset \xRightarrow[\beta_{m^*} g(\mathbf{X},t^k)]{\beta_m f(\mathbf{X})} \mathcal{M}}_{\text{Transcription of DNA to mRNA}}, &\quad 
		\underbrace{\mathcal{M} \xrightarrow{\beta_p} \mathcal{M} + \mathcal{P}}_{\text{Translation of mRNA to Protein}},\\ \quad
		\underbrace{\mathcal{M} \xrightarrow{\gamma_m} \emptyset}_{\text{Degradation of mRNA}}, &\text{ and } 
		\underbrace{\mathcal{P} \xrightarrow{\gamma_p} \emptyset}_{\text{Degradation of Protein}},
	\end{split}
	 \label{eq:dautoIn}
\end{equation}
where chemical species $\mathcal{M}$ and $\mathcal{P}$, respectively, represent the mRNA and protein molecules. For the delay reaction for mRNA transcription, $\beta_m>0$ is the maximum initiation rate and $\beta_{m^*}>0$ is the maximum completion rate. For the remaining instantaneous reactions, $\beta_p$ is the translation rate, $\gamma_m$ is the mRNA degradation rate, and $\gamma_p$ is the protein degradation rate. Based on the formalism from sections \ref{subsec: 2.1 NM-BCRN} and \ref{subsec: 2.1.2 slow reactions} we have reaction sets $\mathcal{R}_D = \{1\}$ and $\mathcal{R}_I = \{2,3,4\}$, the system state at time $t> t_0$ is $\mathbf{X}(t) = [M(t),P(t),M^*(t)]^{\text{T}}$ with the copy numbers given by $M(t)\geq 0$ for mRNA, $P(t)\geq0$ for proteins, and $M^*(t) \geq 0$ for incomplete mRNA transcriptions. The stoichiometric matrices and elementary vectors described for \eqref{eq:kurtz_delay} are
\begin{equation*}
	\nu^- = \begin{bmatrix}
		0 &1 &1 &0\\
		0 &0 &0 &1\\
		0 &0 &0 &0
	\end{bmatrix}, 
	\nu^+ = \begin{bmatrix}
		1 &1 &0 &0\\
		0 &1 &0 &0\\
		0 &0 &0 &0
	\end{bmatrix},
	\nu = \begin{bmatrix}
		1 &0 &-1 &0\\
		0 &1 &0 & -1\\
		0 &0 &0 &0
	\end{bmatrix}, \text{ and } \eta_1 = \begin{bmatrix}
		0\\
		0\\
		1
	\end{bmatrix},
\end{equation*} 
and the propensity functions are
\begin{align}
	\lambda_1(\mathbf{X}(t)) &= \beta_m f(\mathbf{X}(t)), \label{eq:Minit}\\
	\lambda_1^*(\mathbf{X}(t),t,t_{0,1}^{*,k}) &= \beta_{m^*} g(\mathbf{X}(t),t-t_{0,1}^{*,k}), \label{eq:Mcomp}\\
	\lambda_2(\mathbf{X}(t)) &= \beta_p M(t), \notag \\
	\lambda_3(\mathbf{X}(t)) &= \gamma_m M(t),\notag \\
	\lambda_4(\mathbf{X}(t)) &= \gamma_p P(t).\notag
\end{align}
The mRNA transcription propensities for initiation (\eqref{eq:Minit}) and completion (\eqref{eq:Mcomp}) are both regulated by Hill functions. For the transcription initiation, the Hill function depends on the protein copy numbers $P(t)$ and capture the auto-inhibition mechanism,
\begin{equation}
	f\left(\mathbf{X}(t)\right) = \frac{P_{a}^h}{P_{a}^h+P(t)^h},\label{eq:autoIn}
\end{equation}
where $h > 0$ is the Hill coefficient and $P_a$ is a constant such that if $P(t)= P_a$ then $f\left(\mathbf{X}(t)\right) = 1/2$. We note that $f\left(\mathbf{X}(t)\right) \in (0,1)$, since  $f\left(\mathbf{X}(t)\right) \to 0$ as $P(t) \to \infty$, and $f\left(\mathbf{X}(t)\right) \to 1$ as $P(t) \to 0$. In the case of transcription completion, the Hill function depends on the mRNA copy numbers, $M(t)$ and $M^*(t)$, and the internal time $t_1^k = t - t_{0,1}^{*,k}$ to capture the delay effect and finite transcription resources, 
\begin{align}
	g\left(\mathbf{X}(t), t^k\right) = \frac{\alpha M_a^\alpha (\beta_{m^*} t^k)^{\alpha-1}}{M_a^\alpha + v(M(t)+M^*(t))^{\alpha}}. \label{eqn: delay prop}
\end{align}
Here, $M_a$ and $v$ are constants such that if $M(t) + M^*(t) = M_a$ then $g(\mathbf{X}(t),t_1^k) = \alpha (\beta_{m^*}t^k)^{\alpha-1}/(1+v)$ and the shape parameter, $\alpha > 0$, is the only parameter that affects the existence of the non-Markovian delay as it leads to a Weibull distribution for the waiting time. For fixed internal time $t^k_1$ we have that $g(\mathbf{X}(t),t_1^k) \in (0,\alpha(\beta_{m^*}t^k)^{\alpha-1})$ using similar arguments to those used for \eqref{eq:autoIn}. Note that if $\alpha = 1$, the dependence on the internal time is removed and the system becomes Markovian. However, when $\alpha > 1$, the completion event becomes more likely to occur as the internal time increases. Conversely, when $\alpha < 1$, the completion event becomes less likely to occur as the internal time increases and can lead to stalling of the transcription process as $g(\mathbf{X}(t),t^k) \to 0$ as the slowing completion processes build up (e.g., $M^* \to \infty$).

\subsection{Example 2: A non-Markovian epidemic model}
Our second example is a non-Markovian extension to the classic Markovian SIS epidemic model, typically used to model the spread of an infection that does not induce any form of immunity upon recovery. Our non-Markovian network is given by
\begin{equation}
	\underbrace{\mathcal{S} \xRightarrow[\beta_{I^*} g_I^*(\mathbf{X},t_I^k)]{\beta_R^* g_R^*(\mathbf{X},t_R^k)} \mathcal{I}}_{\text{Infection process}}, \text{ and }
	\underbrace{\mathcal{I} \xRightarrow[\beta_R^* g_R^*(\mathbf{X},t_R^k)]{\beta_{I^*} g_I^*(\mathbf{X},t_I^k)} \mathcal{S}}_{\text{Recovery Process}}. \label{eq:SIS_model}
\end{equation}
where the species $\mathcal{S}$ and $\mathcal{I}$, respectively, represent the susceptible and infectious individuals in the population of size $P$. The rate $\beta_I^* >0$ is the maximum completion rate for infection process and the maximum initiation rate for the recovery process. Similarly, $\beta_R^* >0$ is the maximum completion rate for recovery process and the maximum initiation rate for the infection process. Here, both reactions are delay type with $\mathcal{R}_D = \{1,2\}$ and $\mathcal{R}_I = \emptyset$. The system state at time $t > t_0$ is given by $\mathbf{X}(t) = [S(t),I(t),S^*(t),I^*(t)]$ with the total populaton given, $S(t) \geq 0$ to susceptibles, $I(t) \geq 0$ for infectious, $S^*(t) \geq 0$ for infectious that are recovering, and $I^*(t)$ for susceptibles becoming infectious. Due to the form of the two reactions in \eqref{eq:SIS_model}, the completion event for the recovery reaction is the initiation event for the infection reaction and vice-versa. As a result, beyond the initial condition, the population is effectively split between the delay states $S^*(t)$ and $I^*(t)$. 

The stoichiometric matrices are given by  
\begin{equation*}
	\nu^- = \begin{bmatrix}
		1 &0\\
		0 &1\\
		0 &0\\
		0 &0
	\end{bmatrix}, 
	\nu^+ = \begin{bmatrix}
		0 &1\\
		1 &0\\
		0 &0\\
		0 &0
	\end{bmatrix},
	\nu = \begin{bmatrix}
		-1 &1\\
		1 &-1\\
		0 &0\\
		0 &0
	\end{bmatrix}, \eta_1 = \begin{bmatrix}
		0\\
		0\\
		0\\
		1
	\end{bmatrix}, \text{ and } \eta_2 = \begin{bmatrix}
	0\\
	0\\
	1\\
	0
	\end{bmatrix}.
\end{equation*}
To completely define the propensities, we denote the internal time for the $k$th recovery and infection completion process, respectively, as $t_R^k = t - t_{0,R}^{*,k}$ and $t_R^I = t - t_{0,I}^{*,k}$. Here, $t_{0,R}^{*,1}, t_{0,R}^{*,2}, \ldots$, are initiation times for recovery reactions arising from infection completion events. Similarly, $t_{0,I}^{*,1}, t_{0,I}^{*,2}, \ldots$, are initiation times for infection reactions arising from recovery completion events.  The propensity functions are, 
\begin{align}
	\lambda_1(\mathbf{X}(t),t,t_{0,R}^{*,k}) = \lambda^*_2(\mathbf{X}(t),t,t_{0,R}^{*,k}) &= \beta_R^* g_R^*(\mathbf{X}(t), t - t_{0,R}^{*,k}), \label{eq:recov}\\
	\lambda^*_1(\mathbf{X}(t),t,t_{0,I}^{*,k}) = \lambda_2(\mathbf{X}(t),t,t_{0,I}^{*,k}) &= \beta_I^* g_I^*(\mathbf{X}(t), t - t_{0,I}^{*,k}), \label{eq:infect},
\end{align}
where $g_R^*(\mathbf{X}(t), t_R^k) = \alpha_R (\beta_R^* t_R^k)^{\alpha_R - 1}$ and $g_I^*(\mathbf{X}(t), t_I^k) = \alpha_I (\beta_I^* t_I^k)^{\alpha_I - 1}(I(t) + S^*(t))/P$ with the total population being $P = S(t) + I(t) + S^*(t) + I^*(t)$, which is a conserved constant quantity. Just as in the delayed gene regulation model, the shape parameters, $\alpha_R > 0$ and $\alpha_I > 0$, correspond to non-Markovian delay behaviour with Weibull completion times, and in the special case that $\alpha_R = \alpha_I = 1$ then the system becomes Markovian.

\subsection{Simulation results}
\label{subsec: 3.2 Forwards Results}
To evaluate the practical performance of our simulation algorithms, we begin by examining sample paths generated from the delayed gene regulation model (\eqref{eq:autoIn}--(\ref{eqn: delay prop})). We generate independent stochastic simulations of this model using both the exact nM-NRM (Algorithm~\ref{alg: nm-nrm}) and the approximate nM-TLM (Algorithm~\ref{alg: nm-tl}). For all auto-inhibition model simulations in these results we use the parameter values $\beta_m = 10$, $\beta_{m^*} = 0.175$, $\beta_p = 1$, $\gamma_m = 0.08$, $\gamma_p =0.05$, $\alpha = 2.5$, $M_a = 10$, $v = 0.5$, $h = 1.5$, and $P_a = 5$, and the initial conditions are set to $\mathbf{X}_0 = [0,0,0]$~\cite{Boguna2014}.

\begin{figure}[h]
	\centering
	\includegraphics[width=0.6\textwidth]{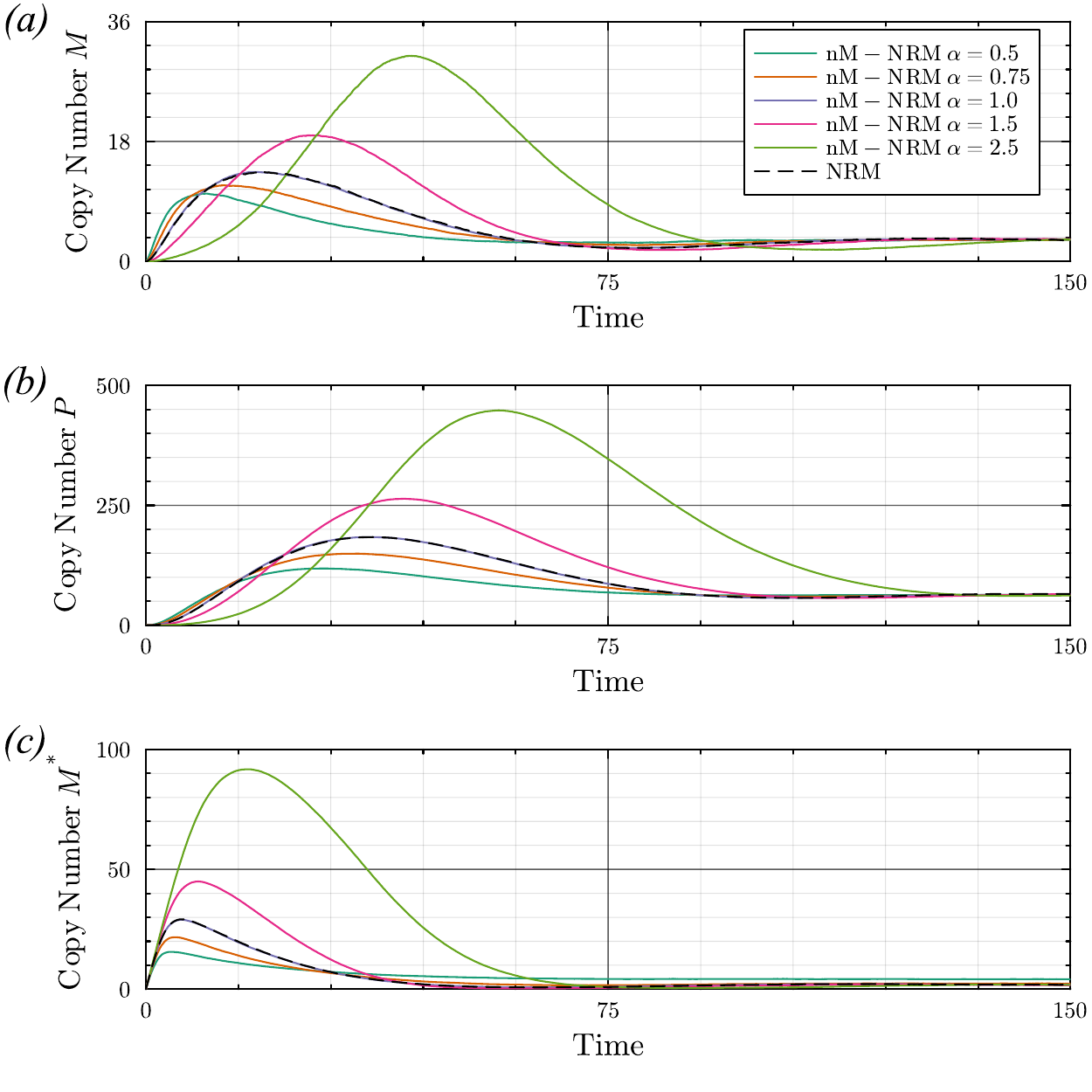}
	\caption{Demonstration of convergence of the non-Markovian setting to the Markovian setting as $\alpha \to 1$. Expected samples paths, computed using $n = 4$,$000$ simulations, are shown for: (a) mRNA molecules $M$, (b) Protein molecules $P$; and (c) progressing transcriptions $M^*$. There is excellent agreement between the $\alpha = 1$ case (purple solid line) and the standard NRM (black dashed line).}
	\label{fig:alpha_conv}
\end{figure}

By inspction of the auto-inhibition model (\eqref{eq:dautoIn}), we see that in the limit as $\alpha \to 1$, the system becomes Markovian as the function $g^*(\cdot)$ loses its time dependence. As a result, our nM-NRM approach should reduce precisely to the Markovian NRM~\cite{Anderson2007,Gibson2000} and give equivalent results to the standard Markovian GDM~\cite{Gillespie1976}. To confirm this we generate MC averaged simulation paths, using $n = 4$,$000$ simulations, for several delay shape parameter values $\alpha = 0.5, 1.0, 1.5, 2.5$. We then compare against the same average generated from the Markovian GDM by removing the time dependence on the transcription completion event. As shown in Figure~\ref{fig:alpha_conv}, we see that there is excellent agreement between our nM-NRM for $\alpha =1$ and the Markovian GDM. Furthermore we note the effect of the history on the dynamics decreases as $\alpha \to 1$. 

\begin{figure}[h]
	\centering
	\includegraphics[width=\textwidth]{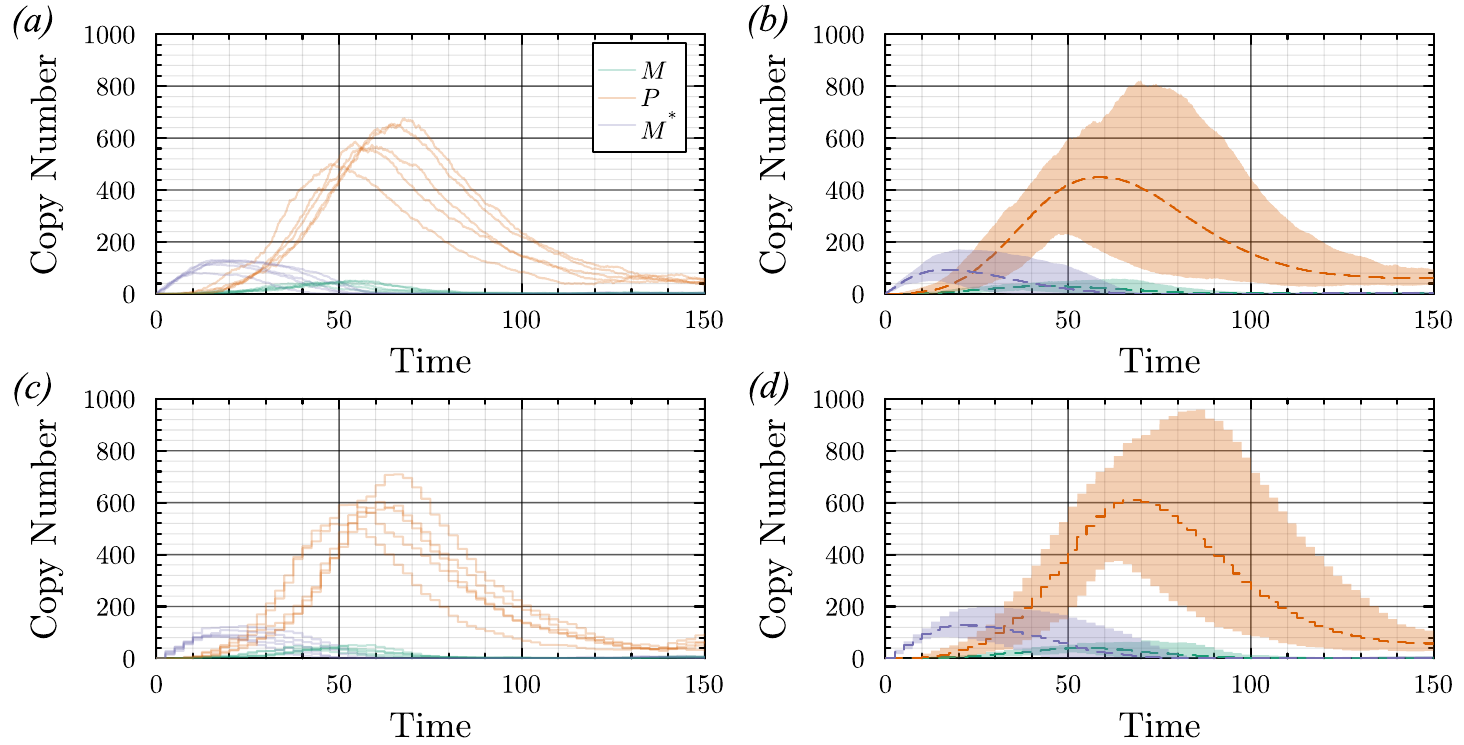}
	\caption{Example realisations of the gene regulation with delayed autoinhibition model (\eqref{eq:autoIn}--(\ref{eqn: delay prop})) using the exact nM-NRM (a)--(b) and approximate nM-TLM with time step $\tau = 2.5$ (c)--(d). $n = 5$ independent paths are shown for (a) nM-NRM and (c) nM-TLM. Path distributions are shown using $n =100$ realisations for (b) nM-NRM and (d) nM-TLM with solid dashed lines indicating mean copy numbers. Model parameters are $\beta_m = 10$, $\beta_{m^*} = 0.175$, $\beta_p = 1$, $\gamma_m = 0.08$, $\gamma_p = 0.05$, $\alpha = 2.5$, $M_a = 10$, $v = 0.5$, $P_a = 5$, and $h = 1.5$.}
	\label{fig: paths}
\end{figure}

Figure~\ref{fig: paths} presents example sample paths from the exact nM-NRM (Algorithm~\ref{alg: nm-nrm}; Figure~\ref{fig: paths}(a)--(b)) and approximate nM-TLM (Algorithm~\ref{alg: nm-tl}; Figure~\ref{fig: paths}(c)--(d)) showing the evolution of mRNA $M(t)$, proteins $P(t)$, and progressing mRNA transcriptions $M^*(t)$. Using a small number of sample paths ($n = 5$) we can observe qualitatively similar behaviour between exact realisations (Figure~\ref{fig: paths}(a)) and their approximate counterparts (Figure~\ref{fig: paths}(c)). In both cases, we observe the expected effect of the auto-inhibition (\eqref{eq:autoIn}) with increase then decline in protein copy numbers, eventually leading to smaller oscillations. However, as expected path-wise numerical error can be observed due to the nM-TLM discretisation $\tau = 2.5$ (Figure~\ref{fig: paths}(c)) compared with exact sample paths (Figure~\ref{fig: paths}(a)). This error propagates into the path distributions (Figure~\ref{fig: paths}(b),(d)) estimated with a larger number of sample paths ($n =100$). Here we see that the nM-TLM (Figure~\ref{fig: paths}(d)) is consistently over-estimating the peak of the protein path (mean $P \approx 600$; Figure~\ref{fig: paths}(d)) in comparison to the exact nM-NRM (mean $P \approx 400$; Figure~\ref{fig: paths}(c)). In addition, the peak of the protein in the nM-TLM approximation is occurring at a later time ( mean peaks at $ t \approx 75$; Figure~\ref{fig: paths}(d)) compared with the exact nM-NRM (mean peaks at $ t \approx 55$; Figure~\ref{fig: paths}(b)). 

\begin{figure}[H]
	\centering
	\includegraphics[width=\textwidth]{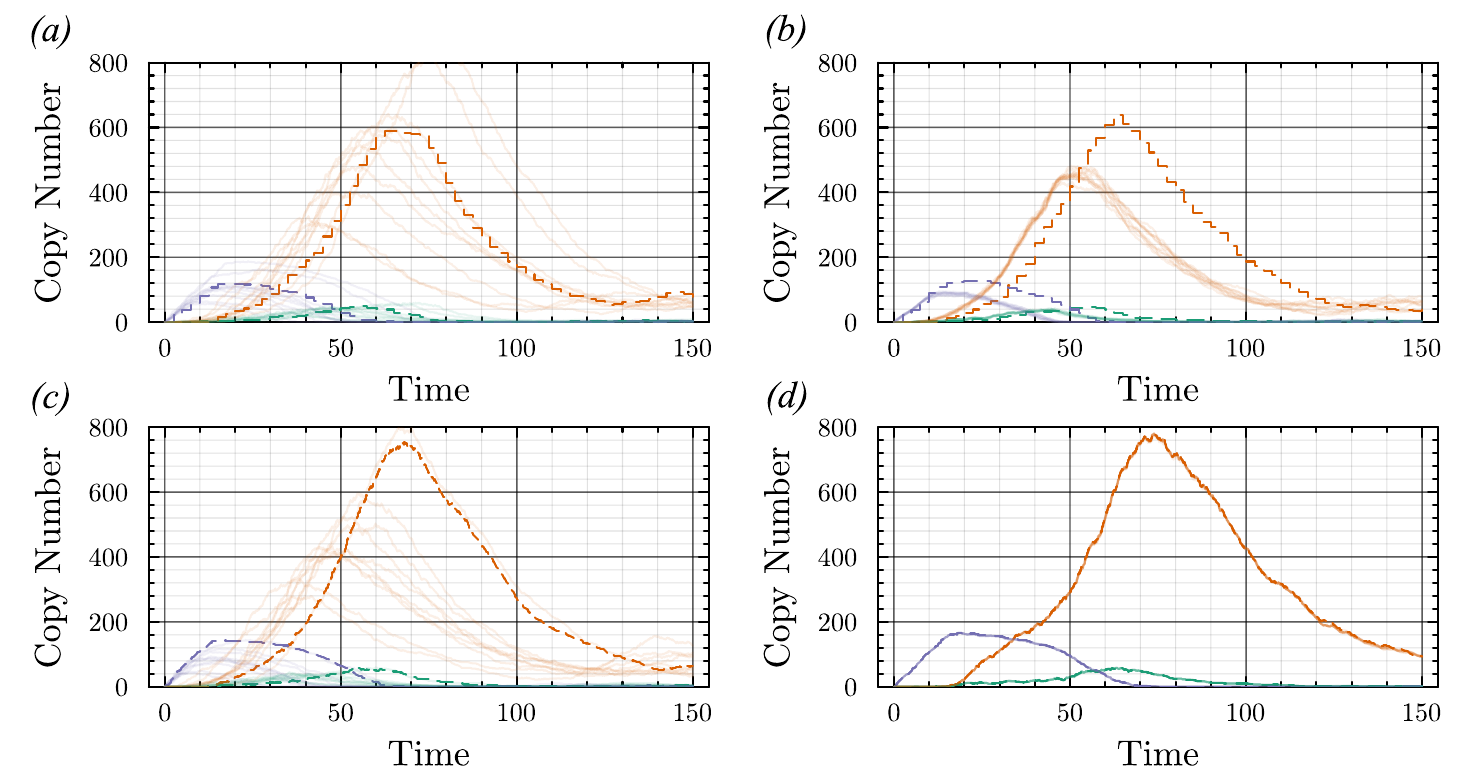}
	\caption{Demonstration of the coupling scheme (Appendix~\ref{app:coupled-nM-NRM}) for nM-NRM realisations (solid lines) that are correlated to a given nM-TLm realisation (dashed lines). Each panel shows $n = 10$ independent nM-NRM simulations  that are: (a) uncoupled and independent of the nM-TLM simulation with $\tau =2.5$; (b) coupled and correlated to the nM-TLM simulation with $\tau =2.5$; (c) uncoupled and independent of the nM-TLM simulation with $\tau =0.05$; and (d) coupled and correlated to the nM-TLM simulation with $\tau=0.05$. Model parameters are $\beta_m = 10$, $\beta_{m^*} = 0.175$, $\beta_p = 1$, $\gamma_m = 0.08$, $\gamma_p = 0.05$, $\alpha = 2.5$, $M_a = 10$, $v = 0.5$, $P_a = 5$, and $h = 1.5$.}
	\label{fig:paths-coupled}
\end{figure}

We also demonstrate the correctness of our coupling scheme. To do this we generate nM-TLM realisations using Algorithm~\ref{alg: nm-tl} and generate $n = 10$ coupled exact nM-NRM realisations (Appendix~\ref{app:coupled-nM-NRM}). This is then compared against $n=10$ independent realisations of the exact nM-NRM using Algorithm~\ref{alg: nm-nrm}. The results, shown in Figure~\ref{fig:paths-coupled}, give a comparison of the independent (Figure~\ref{fig:paths-coupled}(a),(c)) and coupled (Figure~\ref{fig:paths-coupled}(b),(d)) cases for two different step sizes, $\tau = 2.5$ (Figure~\ref{fig:paths-coupled}(a)--(b)) and $\tau = 0.05$ (Figure~\ref{fig:paths-coupled}(c)--(d)). We note a substantial reduction in variance in the exact realisations due to the coupling scheme (compare Figure~\ref{fig:paths-coupled}(a) with Figure~\ref{fig:paths-coupled}(b) and Figure~\ref{fig:paths-coupled}(c) with Figure~\ref{fig:paths-coupled}(d)). The coupled simulations also clearly demonstrate the approximation error due to the discretisation $\tau= 2.5$ (Figure~\ref{fig:paths-coupled}(b)) and convergence of nM-TLM to the nM-NRM as $\tau \to 0$ (Figure~\ref{fig:paths-coupled}(d)). Note that the $n =  10$ exact simulations in Figure~\ref{fig:paths-coupled}(a) are independent simulations to the $n = 5$ samples in Figure~\ref{fig: paths}(a).

\begin{figure}[H]
	\centering
	\includegraphics[width=0.45\linewidth]{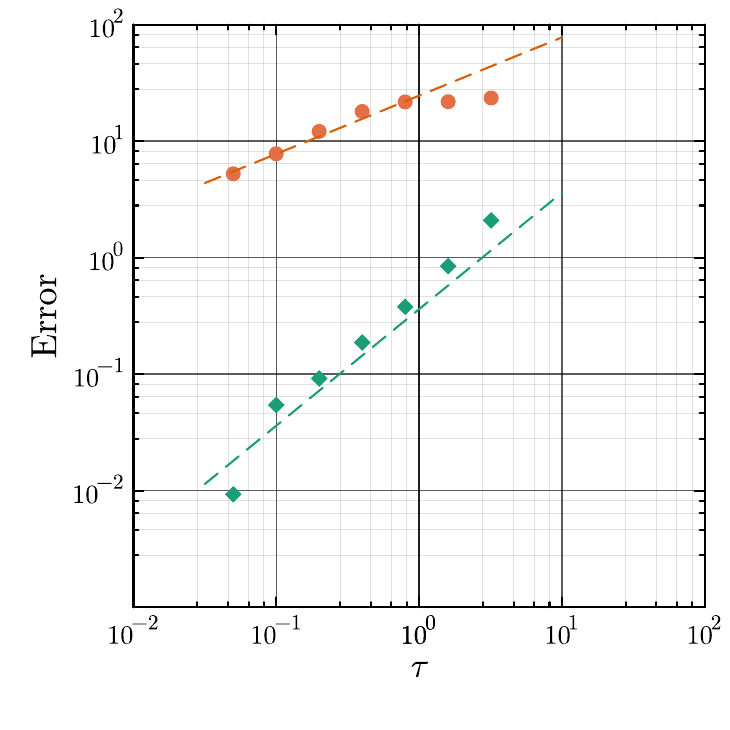}
	\caption{Empirical convergence of the nM-TLM under weak error (green diamonds) and strong error (orange circles). Lines corresponding to a weak convergence rate of $\mathcal{O}(\tau)$ (green dashed line) and a strong convergence rate of $\mathcal{O}(\tau^{1/2})$ (orange dashed line) are show for reference.}
	\label{fig:conv}
\end{figure}

Using coupled simulations pairs $(\mathbf{X}(T), \mathbf{Z}(T))$, we can numerically estimate the error rates of our nM-TLM. Specifically we consider the weak error,
\begin{equation}
	E_{\text{weak}} = \left|\expect{\mathbf{X}(T)} - \expect{\mathbf{Z}(T)} \right|,
	\label{eq:weakerr}
\end{equation}  
and the strong error,
\begin{equation}
	E_{\text{strong}} = \expect{\left| \mathbf{X}(T) - \mathbf{Z}(T)\right|}.
	\label{eq:strongerr}
\end{equation}
Using direct MC with $n = 120,000$ simulations and $T = 300$, we estimate (\ref{eq:weakerr}) and (\ref{eq:strongerr}) for a range of values for $\tau$. This enables us to quantify empirical convergence rates in both the weak and strong sense. These empirical rates are shown in Figure~\ref{fig:conv} along with reference lines for typical weak and strong orders of convergence for a first order scheme such as standard Markovian $\tau$-leaping method and the Euler-Maruyama scheme for stochastic differential equations~\cite{Gillespie2001,Higham2008}. That is, order $\mathcal{O}(\tau)$ for the weak convergence rate and order $\mathcal{O}(\tau^{1/2})$ for the strong convergence rate. We observe good agreement between the empirical rates and the expected theoretical rates.

Finally, we demonstrate that this coupling scheme is also effective for driving computational gains using the unbiased MLMC estimator in \eqref{eq:mlmc_est}. Here we use the non-Markovian SIS model (\eqref{eq:SIS_model}), with parameters $\alpha_R = 1.8$, $\beta_R^*= 1.5$, $\alpha_I = 1.3$, $\beta_I^* = 2.0$, and initial conditions $S^*(t_0) = 1$,$583$, $I^*(t_0)  = 417$, $S(t_0) = I(t_0) = 0$~\cite{Boguna2014}. We apply the unbiased MLMC to estimate $\expect{I^*(T)}$ at $T = 3$, for a range of discretisations $\tau \in \{0.01,0.05, 0.1, 0.2, 0.5,\}$. We take a range of samples sizes (up to $N = 4$,$000$) for direct MC estimators using the exact nM-NRM (Algorithm~\ref{alg: nm-nrm}), and apply the MLMC estimator (\eqref{eq:mlmc_est}) with optimal $N_0$ and $N_1$ for each value of $\tau$ for equivalent target mean-squared error (MSE) for the direct MC. For each sample configuration (samples size and $\tau$), we estimate the MSE and total computational cost (that is, the runtime). 
\begin{figure}[h]
	\centering
	\includegraphics[width=0.7\textwidth]{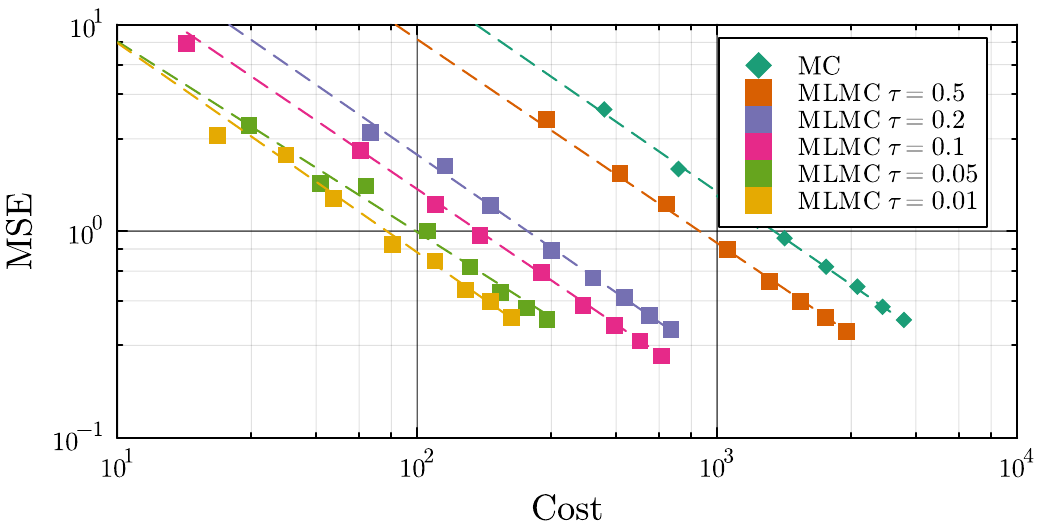}
	\caption{Illustration of the MSE of the estimate of $\expect{I^*(T)}$ as a function of computational cost (that is, the total runtime in seconds) for the direct MC estimator compared with the MLMC estimator with different nM-TLM discretisations $\tau$. The target expectation is $\expect{I^*(T)}$ under the non-Markovian SIS model with $T = 3$, $\alpha_R = 1.8$, $\beta_R^*= 1.5$, $\alpha_I = 1.3$, $\beta_I^* = 2.0$, and initial conditions $S^*(t_0) = 1$,$583$, $I^*(t_0)  = 417$, $S(t_0) = I(t_0) = 0$.}
	\label{fig:mlmc-conv}
\end{figure}

Results in Figure~\ref{fig:mlmc-conv} show that we can achieve the substantial performance improvements of an order of magniture (in terms of MSE for equivalent cost to direct MC). An important observation here is that smaller $\tau$ can be more computationally efficient than larger $\tau$ as the effect of the coupling is stronger in this setting.

\subsection{Inference results}
\label{subsec: Backwards Results}

Given the simulation results (Figures~\ref{fig: paths}--\ref{fig:conv}), we have the necessary components to implement the multifidelity ABC inference scheme (Algorithm~\ref{alg: MF-ABC}). We generate a synthetic dataset using nM-NRM for the delayed auto-inhibition model (Algorithm~\ref{alg: nm-nrm}) to replicate an observed experimental time series of gene transcription. We then evaluate the efficiency of the multifidelity ABC scheme that is accelerated using the nM-TLM (Algorithm~\ref{alg: nm-tl}) with exact coupling (Appendix~\ref{app:coupled-nM-NRM}).

\begin{figure}[H]
	\centering
	\includegraphics[width=0.8\textwidth]{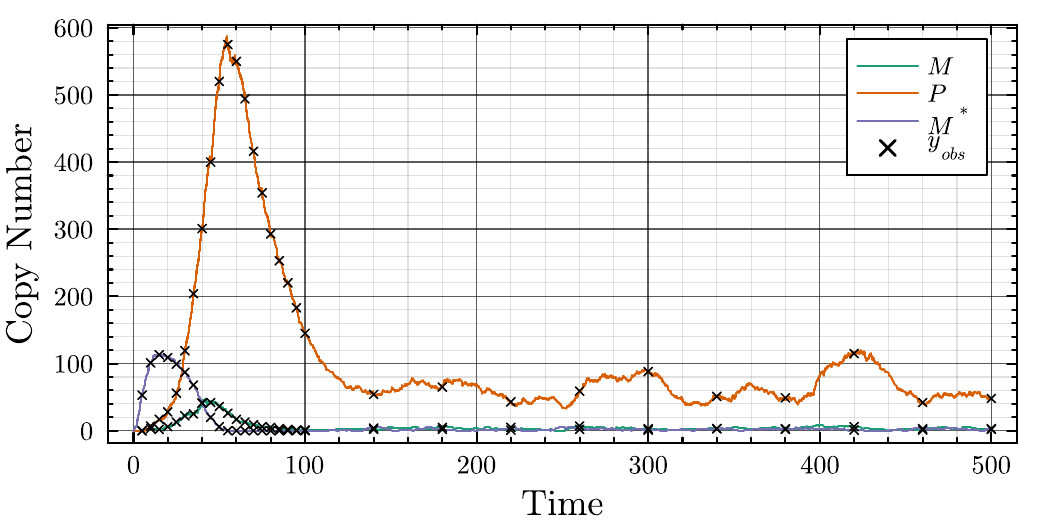}
	\caption{Synthetic data generated from an exact realisation of the gene transcription with delayed auto-inhibition model (\eqref{eq:autoIn}--(\ref{eqn: delay prop})). The full sample path for mRNA $M(t)$ (green line), proteins $P(t)$ (orange line), and incomplete transcriptions $M^*(t)$ (purple line) are shown along with the discrete observations contributing to the data $y_{obs}$ (black crosses). Here the model parameters are $\beta_m = 10$, $\beta_{m^*} = 0.175$, $\beta_p = 1$, $\gamma_m = 0.08$, $\gamma_p =0.05$, $\alpha = 2.5$, $M_a = 10$, $v = 0.5$, $h = 1.5$, and $P_a = 5$. The initial condition is $M(t_0) = P(t_0) = M^*(t_0) = 0$.}
	\label{fig: Yobs}
\end{figure}

The observed data is denoted as $y_{\text{obs}} = [\mathbf{X}(t_1), \mathbf{X}(t_2), \ldots, \mathbf{X}(t_n)]$ where $\mathbf{X}(t_i) = [M(t_i),P(t_i),M^*(t_i)]^{\text{T}}$ with the observations times $t_i = t_0 + 5i$ for $i \leq 20$ and $t_i = t0 + 20i$ for $i > 20$. The data is generated using a single exact realisation of the delayed auto-inhibition model (\eqref{eq:dautoIn}) using the nM-NRM (Algorithm~\ref{alg: nm-nrm}) with initial condition $M(t_0) = P(t_0) = M^*(t_0) = 0$ and parameter values $\beta_m = 10$, $\beta_{m^*} = 0.175$, $\beta_p = 1$, $\gamma_m = 0.08$, $\gamma_p =0.05$, $\alpha = 2.5$, $M_a = 10$, $v = 0.5$, $h = 1.5$, and $P_a = 5$. This synthetic data and the full underlying realisation is shown in Figure~\ref{fig: Yobs}. Note the observations times start at a higher resolution to capture the initial peak in the gene expression. 

For the inference problem we consider the mRNA maximum completion rate, $\beta_{m^*}$, the protein production rate parameters, $\beta_p$, and the transcription completion time shape parameter, $\alpha$, to be unknown and to be inferred from the data, $y_{\text{obs}}$. Specifically we consider the problem of estimating the posterior mean,
\begin{equation*}
	\condexpect{\theta}{y_{\text{obs}}} = \int_\Theta \theta \condpdf{\theta}{y_{\text{obs}}}\, \text{d}\theta,
\end{equation*} 
where $\theta = [\beta_{m^*},\beta_p,\alpha]^{\text{T}}$. The priors used are independent uniform priors, $\beta_{m^*} \sim \mathcal{U}(0,0.5)$, $\beta_p \sim \mathcal{U}(0,5)$ and $\alpha \sim \mathcal{U}(0,10)$. We choose uniform priors for simplicity and highlight that our methods are not restricted to uniform priors. We use this problem set-up to demonstrate the efficiency gains we obtain through the multifidelity ABC method.

We compare the ABC accept-reject scheme (Algorithm~\ref{alg: ABC rej}) with exact nM-NRM simulations (Algorithm~\ref{alg: nm-nrm}) against the multifidelity ABC accept-reject of Prescott and Baker~\cite{Prescott2020} that uses the nM-TLM (Algorithm~\ref{alg: nm-tl}) for low-fidelity simulations and the coupling scheme (Appendix~\ref{app:coupled-nM-NRM}) for the correction terms (\eqref{eq:mf_w}). In both cases the ABC approximation is taken with respect to the Euclidean discrepancy metric on the full data, that is
\begin{equation*}
	\rho(y_{\text{sim}},y_{\text{obs}}) = \left(\sum_{i=1}^n \| \mathbf{X}_{\textrm{sim}}(t_i) - \mathbf{X}_{\textrm{obs}}(t_i) \|_2^2\right)^{1/2}
\end{equation*}
where $\| \cdot \|_2$ is the vector $2$-norm. The acceptance threshold is taken as $\varepsilon = 1200$. 

To implement the multifidelity ABC scheme we need to optimise the continuation probability function $\mu(\tilde{y}_{\text{sim}})$ given in \eqref{eq:mf_mu}. Following the tuning process outlined be Prescott and Baker~\cite{Prescott2020} we initially set $\mu_a = 1$ and $\mu_r = 1$ and preform a relatively small number of warm-up samples ($N = 20,000$) to refine these continuation probabilities according to the optimality condition set out in Appendix~\ref{app:contprob}. We apply this tuning process for $\tau= 0.68$ and $\tau = 2.31$. We note that our exact coupling scheme (Appendix~\ref{app:coupled-nM-NRM}) is essential for achieving high performance in multifidelity ABC. 





\begin{figure}[H]
	\centering
	\includegraphics[width=\textwidth]{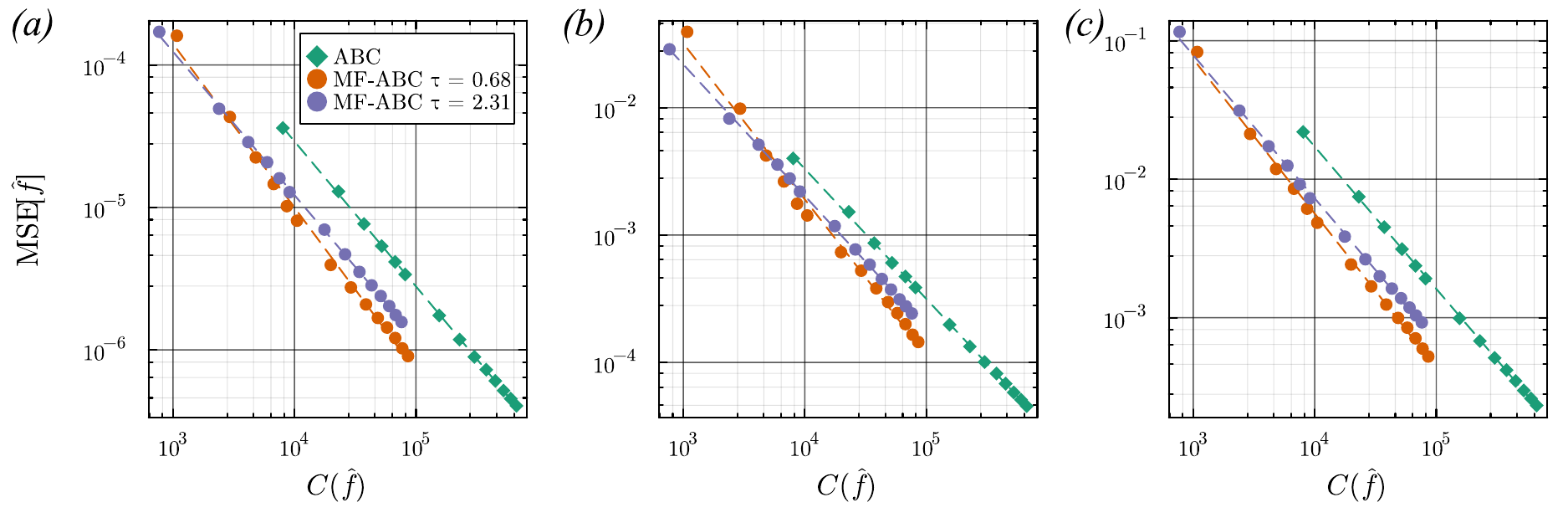}
	\caption{Comparison of convergence rate of posterior mean estimator MSE as a function of computational cost (that is, total runtime in seconds) for: (a) mRNA transcription initiation rate $\beta_m$; (b) protein translation rate $\beta_p$; (c) transcription completion time shape parameter $\alpha$. Results are shown for standard ABC methods (green diamonds), MF-ABC with $\tau = 0.68$ (orange circles), and MF-ABC with $\tau = 2.31$ (purple squares). Regression lines are also shown (dashed lines). The ABC discrepancy threshold used is $\varepsilon = 1200$.}
	\label{fig: 3panelVar}
\end{figure}

We estimate posterior means using the direct ABC rejection sampling and multifidelity ABC sampling for a range of MC sample sizes. Denoting the estimate as $\hat{f}$, in each instance we estimate the MSE of this estimate, $\text{MSE}[\hat{f}]$, and the average computational cost (total runtime) to obtain the estimate, $C(\hat{f})$. Here the MSE is measured against a gold standard ABC estimate with $N = 1,000,000$, and the computational cost is the average runtime to ensure fair comparison as the total number of samples will be smaller for ABC accept-reject than the multifidelity approach of equivalent computational cost. Figure~\ref{fig: 3panelVar} demonstrates the empirical convergence of this variance as a function of computational cost for the mRNA maximum completion rate, $\beta_{m^*}$ (Figure~\ref{fig: 3panelVar}(a)), the protein production rate $\beta_p$ (Figure~\ref{fig: 3panelVar}(b)), and the transcription completion time shape parameter, $\alpha$ (Figure~\ref{fig: 3panelVar}(c)). In each case, the measurements align with a rate of $\text{MSE}[\hat{f}] = \mathcal{O}(C(\hat{f})^{-1})$ as is expected from  the theory in Prescott et al.,~\cite{Prescott2024} for a fixed acceptance threshold, $\varepsilon > 0$.

\begin{figure}[H]
	\centering
	\includegraphics[width=\textwidth]{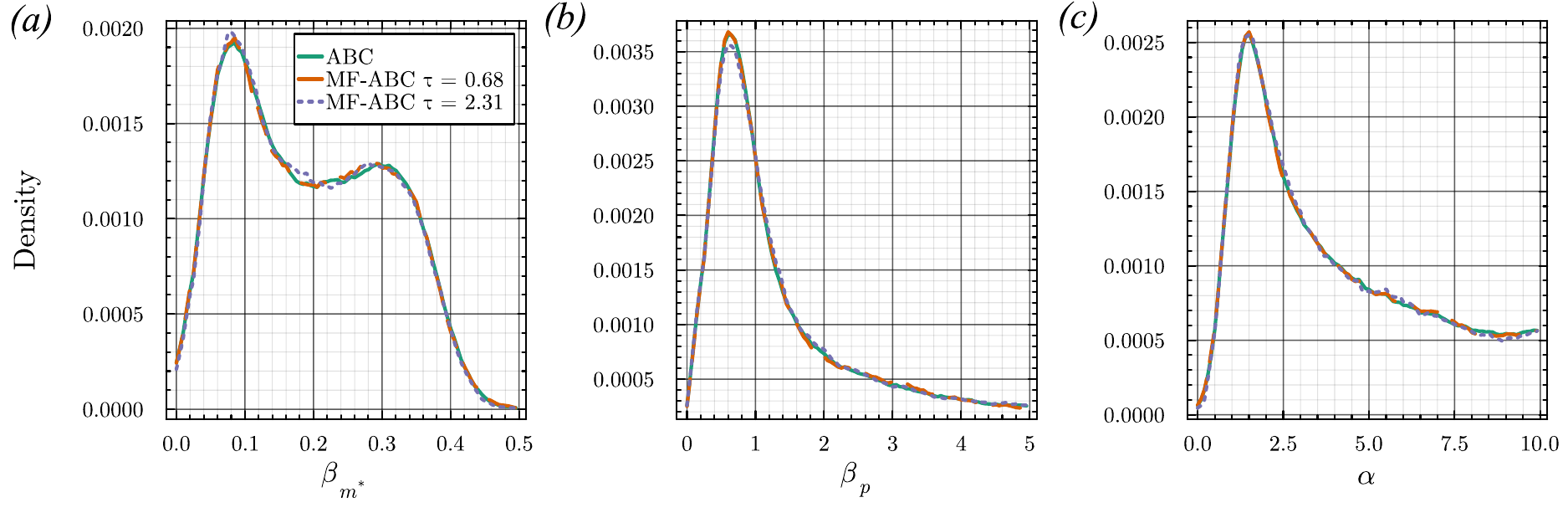}
	\caption{Comparison of posterior marginal probability density estimates obtained for: (a) mRNA transcription initiation rate $\beta_m^*$; (b) protein translation rate $\beta_p$; (c) transcription completion time shape parameter $\alpha$. Results are shown for the gold standard ABC samples (green solid line), MF-ABC with $\tau = 0.68$ (orange dashed line), and MF-ABC with $\tau = 2.31$ (purple dotted line). The MF-ABC densities are obtained with a factor $10\times$ Computational speed-up. The ABC discrepancy threshold used is $\varepsilon = 1200$.}
	\label{fig: 3panelpdf}
\end{figure}

Across all parameters we observe a reduction in estimator MSE for equivalent levels of computation cost (that is, total runtime). The level of improvement depends on the value of $\tau$, just as in the forward problem case. While $\tau = 0.68$ achieves a small improvement over $\tau = 2.31$,  in both cases the speed-up factor is approximately $5\times$ over direct implementation of ABC. These results demonstrate the efficacy of our approximate stochastic simulation and exact coupling schemes can be exploited in to substantially accelerate simulation-based inference approaches though variance reduction of bias correction terms~\cite{Prescott2020,Warne2018}. Beyond this demonstration of the posterior mean estimator, we can also demonstrate accelerated posterior probability density estimation. Figure~\ref{fig: 3panelpdf} shows excellent agreement between the posterior marginal probability densities obtained using the MF-ABC samples used in Figure~\ref{fig: 3panelVar} (each with total cost $C(\hat{f}) \approx 10^{5}$ seconds) and expensive ABC gold standard (with total cost $C(\hat{f}) \approx 10^{6}$ seconds). This accuracy holds across all three unknown parameters, $\beta_m^*, \beta_p$,  and $\alpha$. While such results have been known and demonstrated before in the context of Markov processes~\cite{Prescott2024,Warne2018,Warne2022}, our methods extend the applicability of multifidelity methods to applications involving non-Markovian systems for the first time. 

\FloatBarrier

\section{Discussion}
\label{sec: 4 conc}
In this work, we present a novel extension to the next reaction method to exactly simulate stochastic networks that are non-Markovian due to time dependent propensities, stochastic delays that depend on both time and system state~\cite{Anderson2007,Boguna2014}. This extension leads to a natural first order approximation of the Kurtz random time-change representation to result in an equivalent non-Markovian $\tau$-leaping method~\cite{Gillespie2001}. Finally, we develop an exact coupling scheme based on common Poisson clocks and demonstrate its applicability to variance reduction methods such as MLMC~\cite{Anderson2012,Giles2008} and multifidelity methods~\cite{Prescott2020,Warne2022}.


While we focus on the original multifidelity ABC method of Prescott and Baker~\cite{Prescott2020} and achieve orders of magnitude improvement in efficiency, it is important to highlight the generality of our approach to alternative methods in the simulation-based inference literature. In particular, any variant of multifidelity scheme that fundamentally relies on coupling of exact and approximation simulation pairs can be applied in our setting. This includes MLMC for ABC~\cite{Jasra2019,Warne2018,Warne2022}, multifidelity SMC for ABC~\cite{Prescott2021} and adaptive extension~\cite{Prescott2024}, and state-of-the-art approaches for multifidelity neural posterior estimation~\cite{Hikida2025,Krouglova2025}. In this work, our main contribution is the necessary stochastic simulation tools to enable MLMC and multifidelity approaches to simulation and inference in the non-Markovian setting.

In addition to parameter inference, our methods are also conducive to acceleration of summary statistic estimators for the forward problem following the Markovian work in networks~\cite{Anderson2012,Lester2016,Warne2019} and stochastic differential equations~\cite{Giles2008,Rhee2015}. The potential fo orders of magnitude improvement in simulation time could be of tremendous significance in the setting of whole cell modelling in quantitative biology that are typically extremely computationally expensive and often rely on stochastic simulation for key biochemical processes~\cite{Feig2019,Roberts2011,Stumpf2021,Yeom2021}.

We primarily demonstrate our non-Markovian simulation schemes in the context of biochemical processes. This is largely due to the substantial amount of stochastic modelling and simulation literature that has arisen from the area of chemical physics~\cite{Anderson2007,Cao2004,Gibson2000,Gillespie1976,Gillespie1977,Gillespie1992,Gillespie2001}. However, our methods are not limited to this setting and are general to discrete-state continuous-time non-Markovian processes of the form given in \eqref{eq:kurtz_delay}. As demonstrated by our epidemiology examples using MLMC simulation, our approach is widely applicable to alternative contexts involving non-Markovian stochastic models, such as ecology~\cite{Fleming2014,Williams2007}, physics~\cite{Boguna2014,Vega2017} or queues~\cite{Karthikeyan2023,Maragathasundari2019}. 

Another area of applicability for this work is stochastic simulation following Markovian projection~\cite{Bayer2018,Gyoengy1986}. In the setting of reaction networks, Markovian projection can substantially reduce the dimensionality of the state vector while preserving the marginal path distribution for a selected subset of state dimensions~\cite{BenHammouda2024}. This can be extremely beneficial in partially observed systems as it enables simulation to be performed directly on the observable chemical species without considering any latent variables. The cost of the Markovian projection is that the reduced system can become non-Markovian~\cite{Hammouda2025}. Our methods will enable both exact and approximate simulation in this promising area of dimensionality reduction.

In the setting of optimal configuration of multifidelity simulation-based inference, whether it be based on ABC or alternatives, one open challenge is the automatic tuning of the step size parameter, $\tau$. Importantly, the choice of $\tau$ does not impact bias due to the exact coupling scheme~\cite{Anderson2012,Warne2019}, however, a there is a trade-off between the computational gains of larger $\tau$ and the receiver operator characteristic of the low-fidelity simulator as a predictor for the high-fidelity simulator (Appendix~\ref{app:contprob}). While the coupling scheme enables optimal continuation probability tuning for a fixed $\tau$~\cite{Prescott2020,Prescott2024}, it does not help with adapting $\tau$. One possible way forward could be to exploit the estimated strong convergence rate (Figure~\ref{fig:conv}) to obtain an optimal scaling $\tau$ within the MLMC telescoping summation~\cite{Warne2022}. Other possible solutions could consider a randomised $\tau$ to generate a family of low-fidelity models and extending \eqref{eq:mf_w} accordingly~\cite{Prescott2024}. In practice a reasonable choice of $\tau$ can be obtained using a small numerical experiment using the warm-up step of MF-ABC that optimises the continuation probabilities for a range of candidate values for $\tau$ (e.g., see Warne et al.,~\cite{Warne2022}), with smaller optimal continuation probabilities indicating more efficient sampling. This approach also provides some simple diagnostics that indicate when $\tau$ is too large to be effective (See Prescott and Baker~\cite{Prescott2020} and Prescott et al.,~\cite{Prescott2024}). In particular, if one or both of the optimal continuation probabilities is close to unity for all candidate values for $\tau$, then the a finer approximation is required.

We provide a versatile extension of approximate stochastic simulation schemes with exact coupling mechanisms in the context of discrete-state continuous-time non-Markovian stochastic processes. We demonstrate the computational performance of our methods for both the generation of sample paths from the forward problem, and accelerated simulation-based inference for the inverse problem. We obtain an order of magnitude improvements in computational efficiency which could lead to new possibilities for the practical application of complex non-Markovian processes in quantitative biology and many other fields of computational science.

\section*{Acknowledgments}
The project is supported by the Australian Research Council (ARC). DJW is supported by an ARC Discovery Early Career Researcher Award (DE250100396). DJW acknowledges support from the ARC Centre of Excellence for the Mathematical Analysis of Cellular Systems (MACSYS; CE230100001). Computational resources and services used in this work were provided by the eResearch Office, Queensland University of Technology, Brisbane, Australia.

%
%
\bibliographystyle{plain}

\newpage
\begin{appendices}

\section{Coupling and continuation probabilities}
\label{app:contprob}

\numberwithin{equation}{section}
\numberwithin{figure}{section}
\numberwithin{algorithm}{section}
\numberwithin{table}{section}
\setcounter{equation}{0}
\setcounter{algorithm}{0}
\setcounter{figure}{0}

The theory of multifidelity ABC does not strictly require a coupling mechanism between the low-fidelity, $\tilde{y}_\text{sim}$, and high-fidelity simulators $y_\text{sim}$. However, to obtain meaningful performance improvements coupling schemes are extremely effective. In this appendix, we demonstrate the importance of coupling scheme as it relates to optimal configuration of the multifidelity continuation probability function.

We denote $\tilde{C}$ as the computational cost of generating a low-fidelity simulation and $C$ as the computational cost of generating a high-fidelity simulation. Furthermore we assume $\tilde{C}/C \ll 1$.  For a continuation probability function, $\mu(\tilde{y}_\text{sim})$, of the form given in Eq. (15) of the main manuscript, Prescott and Baker~\cite{Prescott2020} show that the optimal continuation probabilities are 
\begin{equation}
	\mu_a^* = \left(\frac{\tilde{c} p_{\text{fp}}}{c_{\text{p}} (p_{\text{tp}} - p_{\text{fp}})}\right)^{1/2}, \text{ and } \mu_r^* = \left(\frac{\tilde{c} p_{\text{fn}}}{c_n (p_{\text{tp}} - p_{\text{fp}})}\right)^{1/2},
\end{equation} 
where $\tilde{c} = \mathbb{E}[\tilde{C}]$, is the expected cost of the low-fidelity simulator, $c_{\text{p}} = \condexpect{C}{\rho(\tilde{y}_{\text{sim}},y_{\text{obs}}) \leq \varepsilon}$ is the expected cost of continuing to a high-fidelity simulation following an accepted low-fidelity simulation, $c_{\text{n}} = \condexpect{C}{\rho(\tilde{y}_{\text{sim}},y_{\text{obs}}) > \varepsilon}$ is the expected cost of continuing to a high-fidelity simulation following a rejected low-fidelity simulation, $p_{\text{fp}} = \condprob{\rho(y_{\text{sim}},y_{\text{obs}}) > \varepsilon}{\rho(\tilde{y}_{\text{sim}},y_{\text{obs}}) \leq \varepsilon}$ is the probability of a false positive (accepted low-fidelity and rejected high-fidelity), $p_{\text{tp}} = \condprob{\rho(y_{\text{sim}},y_{\text{obs}}) \leq \varepsilon}{\rho(\tilde{y}_{\text{sim}},y_{\text{obs}}) \leq \varepsilon}$ is the probability of a true positive (accepted low-fidelity and accepted high-fidelity), and $p_{\text{fn}} = \condprob{\rho(y_{\text{sim}},y_{\text{obs}}) \leq \varepsilon}{\rho(\tilde{y}_{\text{sim}},y_{\text{obs}}) > \varepsilon}$ is the probability of a false negative (rejected low-fidelity and accepted high-fidelity).

\begin{figure}[h]
	\centering
	\begin{subfigure}[t]{0.5\textwidth}
		\centering
		\includegraphics[width=\textwidth]{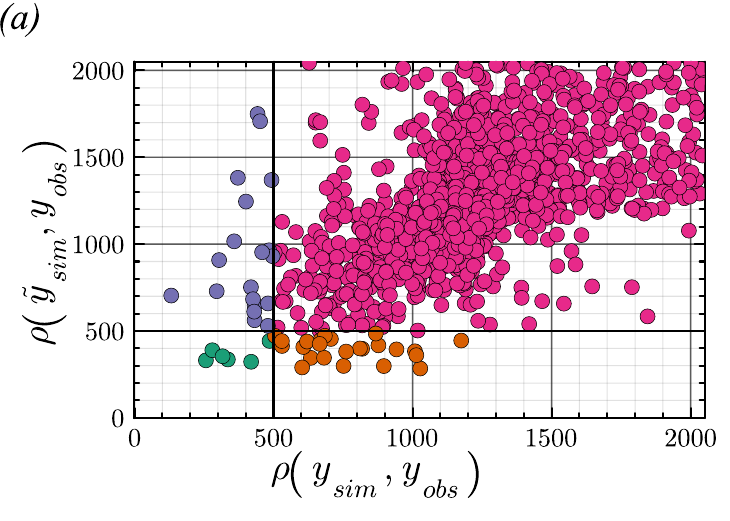}
		\label{subfig: uncor scat 0.25}
	\end{subfigure}%
	~
	\begin{subfigure}[t]{0.5\textwidth}
		\centering
		\includegraphics[width=\textwidth]{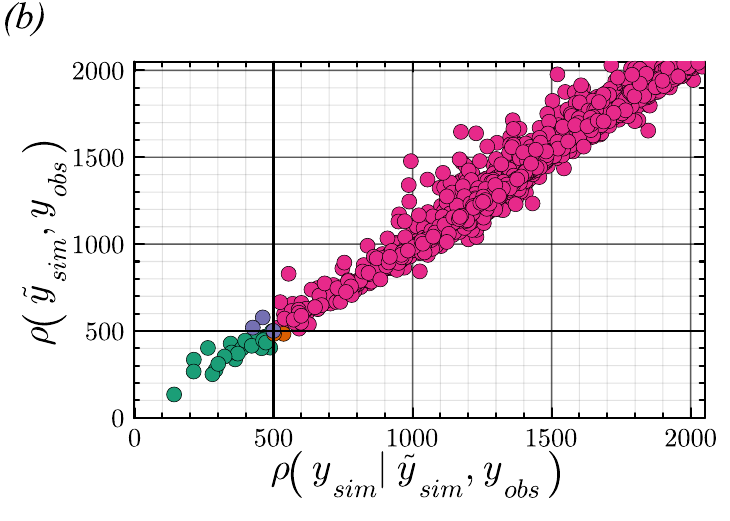}
		\label{subfig: scat 0.25}
	\end{subfigure}%
	\caption{Scatter plots interpreted as confusion matrices of discrepancy measures from the observed data, for (a) uncoupled and (b) coupled sample paths using $\tau = 0.25$. The acceptance threshold of $\varepsilon = 500$ is indicated (sold black lines) to highlight the true positive region (green circles), true negative region (pink circles), false positive region (orange circles) and false negative region (purple circles).}
	\label{fig: cor scats}
\end{figure}

The smaller $\mu_a^*$ and $\mu_r^*$ are, the less frequently high-fidelity simulations are needed to correct for errors accumulated though the use of the low-fidelity simulator, leading to computational gains. In the context of non-Markovian schemes we consider in this paper, we have no control over $c_{\text{p}}$ and $c_{\text{n}}$ as it is based on nM-NRM (Alg.~3), however, $\tilde{c}$ can be reduced by increasing the time step $\tau$ in the nM-TLM (Alg.~4). Doing so will increase the error rate in the low-fidelity accept/reject decision as a predictor for the high-fidelity accept/reject decision.  That is, as $\tau$ increases,  $p_{\text{fp}}$ and $p_{\text{fn}}$ will increase and $p_{\text{tp}}$ will decrease, leading potentially only negligible reduction in $\mu_a^*$ and $\mu_r^*$ below unity. When coupling is introduced using Algs. 5 and 6, the low-fidelity and high-fidelity simulations become positively correlated. This, in turn, leads to improved predictive characteristics with decreased $p_{\text{fp}}$ and $p_{\text{fn}}$ and increased $p_{\text{tp}}$. This effect is visualised in Fig.~\ref{fig: cor scats} for simulation pairs of the discrepancy metric under the low-fidelity and high-fidelity simulation.  

\newpage
\section{Coupling exact and approximate paths}
\label{app:coupled-nM-NRM}

\numberwithin{equation}{section}
\numberwithin{figure}{section}
\numberwithin{algorithm}{section}
\numberwithin{table}{section}
\setcounter{equation}{0}
\setcounter{algorithm}{0}
\setcounter{figure}{0}

Here we develop a coupling scheme to generate sample path pairs $\{(\mathbf{X}(t),\mathbf{Z}(t))\}_{t\geq t_0}$ such that the sample paths are positively correlated for any $t > t_0$. This enables computation cost reductions using MLMC or multifidelity schemes (Alg.~\ref{alg: MF-ABC}). Our approach involves two steps: (i) we perform the completion of Poisson processes using firing counts and interval lengths from a realisation of $\mathbf{Z}(t)$ using nM-TLM (Alg.~\ref{alg: nm-tl}), then (ii) we decompose the delayed reaction channels in the nM-TLM (Alg.~\ref{alg: nm-tl}) to match the exact setting in nM-NRM (Alg.~\ref{alg: nm-nrm}) where copy numbers for all delayed chemical species are at most unity.    

Suppose we simulate $\mathbf{Z}(t)$ for $n$ time steps of length $\tau$ with $t_i = t_0 + i \tau$ for $i = 0, 1, 2, \ldots, n$. At time $t_i$ for a given reaction channel $j \in \mathcal{R}$ (resp. $\kappa \in \mathcal{D}_j$ and $j \in \mathcal{R}_D$ for progressing delay channels), denote $P_{i,j}$ (resp. $P_{i,j}^\kappa$) to be the number of events that occur over the time interval $[t_i, t_{i+1}) = [t_i, t_{i} +\tau)$. We can relate this time interval to equivalent intervals for the unit rate processes, $Y_j(\cdot)$ (resp. $Y^\kappa_j(\cdot)$), via the random time-change formulae (Equations~(\ref{eq:nrt})--(\ref{eq:nrtd})). That is the interval, $[t_{i,j,u}, t_{i+1,j,u})$ (resp. $[t^\kappa_{i,j,u}, t^\kappa_{i+1,j,u})$), over which the unit rate process fires $P_{i,j}$ (resp. $P_{i,j}^\kappa$ ) times, is given by,
\begin{equation}
	t_{i,j,u} = \Lambda_j(t_i; \mathbf{Z}(t_0), t_0), \text{ and }	t_{i+1,j,u} = \Lambda_j(t_{i+1}; \mathbf{Z}(t_0), t_0) = t_{i,j,u} + \Lambda_j(\tau; \mathbf{Z}(t_i), t_i), \label{eq:unit_intervals}
\end{equation}  
and respectively,
\begin{equation}
	t^\kappa_{i,j,u} = \Lambda^*_j(t_i; \mathbf{Z}(t_0), t_0, t_{j,0}^{*,\kappa}), \text{ and }	t^\kappa_{i+1,j,u} = \Lambda^*_j(t_{i+1}; \mathbf{Z}(t_0), t_0, t_{j,0}^{*,\kappa}) = t^\kappa_{i,j,u} + \Lambda^*_j(\tau; \mathbf{Z}(t_i), t_i, t_{j,0}^{*,\kappa}). \label{eq:unit_intervalsd}
\end{equation}  
For a homogeneous Poisson processes, $Y_j(\cdot)$ (resp. $Y_j^\kappa(\cdot)$), the time of an event that occurs within the interval $[t_{i,j,u}, t_{i+1,j,u})$ (resp. $[t^\kappa_{i,j,u}, t^\kappa_{i+1,j,u})$) is a uniformly distributed random variable over the interval. This leads to Alg.~\ref{alg: complete PoisProc} for constructing a sequence of event times for the unit rate Poisson processes in Eq.~\eqref{eq:kurtz_delay} based on the approximate sample path generated through Alg.~\ref{alg: nm-tl}. The resulting sets of event times $\mathcal{T}_j$ for $j \in \mathcal{R}$ can be used in place of sampling the next reaction intervals $r_{j,u} \sim \text{Exp}(1)$ in Alg.~\ref{alg: nm-nrm}, that is the $m$th sample of $r_{j,u}$ can be computed using $r_{j,u}^{(m)} = t_{j,u}^{(m)} - t_{j,u}^{(m-1)}$ where $t_{j,u}^{(m)} = \min{(\mathcal{T}_j \backslash \{t_{j,u}^{(1)},t_{j,u}^{(2)},\ldots,t_{j,u}^{(m-1)} \})}$.
\begin{algorithm}[H]
	\caption{Poisson process completion}
	\begin{algorithmic}[l]
		\Require Event counts, $P_{i,j}$, for $j \in \mathcal{R}$, and $P_{i,j}^\kappa$ for $\kappa = 1, 2, \ldots, K_j$ where $K_j = \sum_{i=1}^n \indd{[1,\infty)}{P_{i,j}}$ and $j \in \mathcal{R}_D$  from a realisation of Alg.~\ref{alg: nm-tl} for times $t_i = t_0 + i\tau$ with $i = 0,1,\ldots,n$,;
		\State Set $\mathcal{T}_j = \{0\}$ as the set of event times for $Y_j(\cdot)$ for $j \in \mathcal{R}$;
		\State Set $\mathcal{T}^\kappa_j = \{0\}$ as the set of event times for $Y^\kappa_j(\cdot)$ for $\kappa = 1, 2, \ldots, K_j$ and $j \in \mathcal{R}_D$;
		\For{$j \in \mathcal{R}$}
		
		\For{$i=1,...,n$}
		\State Compute interval for unit rate process $[t_{i,j,u}, t_{i+1,j,u})$ using Eq.~\eqref{eq:unit_intervals}
		\State Sample event times $t_{j,u}^{(m)} \sim \mathrm{Uniform}(t_{i,j,u},t_{i+1,j,u})$ for $m =1,2, \ldots, P_{i,j}$;
		\State $\mathcal{T}_j \leftarrow \mathcal{T}_j \cup \{t_{j,u}^{(1)},t_{j,u}^{(2)},\ldots,t_{j,u}^{(m)}\}$
		\If{$j \in \mathcal{R}_D$}
		\For{$\kappa = 1,2, \ldots, K_j$}
		\State Compute interval for unit rate process $[t^\kappa_{i,j,u}, t^\kappa_{i+1,j,u})$ using Eq.~\eqref{eq:unit_intervalsd}
		\State Sample event times $t_{j,u}^{\kappa,(m)} \sim \mathrm{Uniform}(t^\kappa_{i,j,u},t^\kappa_{i+1,j,u})$ for $m =1,2, \ldots, P^\kappa_{i,j}$;
		\State $\mathcal{T}^\kappa_j \leftarrow \mathcal{T}^\kappa_j \cup \{t_{j,u}^{\kappa,(1)},t_{j,u}^{\kappa,(2)},\ldots,t_{j,u}^{\kappa,(m)}\}$
		\EndFor
		\EndIf
		\EndFor
		\EndFor
	\end{algorithmic}
	\label{alg: complete PoisProc}
\end{algorithm}

Alg.~\ref{alg: complete PoisProc} is almost all that is needed to implement the nM-NRM using the event times $\mathcal{T}_j$ (resp. $\mathcal{T}_j^\kappa$) for the unit time processes $Y_j(\cdot)$ (resp. $Y_j^\kappa(\cdot)$). However the process $Y_j^\kappa(\cdot)$ constructed here in Alg.~\ref{alg: complete PoisProc} is not identical to the $Y_j^k(\cdot)$ in Eq.~\eqref{eq:kurtz_delay}. This is due to the fact that within the nM-TLM (Alg.~\ref{alg: nm-tl}) groups of delayed chemical species are handled with a single delayed reaction channel. In contrast, nM-NRM (Alg.~\ref{alg: nm-nrm}) handles each initiated delayed reaction with a unique delay chemical species having its own reaction channel. This requires the groups to be disassembled into distinct reaction channels with a single event time.

Following from~Eq.~\eqref{alg: complete PoisProc} for a delayed channel $j \in \mathcal{R}_D$, the set $\mathcal{T}_j$ contains all initiation times, and the sequence of sets $\mathcal{T}_j^1, \mathcal{T}_j^2, \ldots \mathcal{T}_j^{K_j}$ contains all the completion times. The group disassembly proceeds as follows, the first $|\mathcal{T}_j^1|$ initiation events from $\mathcal{T}_j$ will be assigned a completion times from $\mathcal{T}_j^1$ at random uniformly without replacement until $\mathcal{T}_j^1 = \emptyset$, then the next $|\mathcal{T}_j^2|$ initiation events from $\mathcal{T}_j$ will be assigned a completion times from $\mathcal{T}_j^2$ at random uniformly without replacement until $\mathcal{T}_j^2 = \emptyset$, and so on until either $\mathcal{T}_j = \emptyset$ or $\mathcal{T}_j^{K_j} = \emptyset$. Combining the completion of the Poisson process with this disassembly procedure provides a set of event times for the unit time Poisson processes needed for the coupled nM-NRM. This effectively proceeds via implementing Alg.~\ref{alg: nm-nrm} with next event times being determined via the Alg.~\ref{alg: complete PoisProc}. The explicit implementation of this coupled nM-NRM is provided in Appendix~\ref{app:coupled-nM-NRM}. Importantly, the process generates an exact realisation from the non-Markovian biochemical reaction network (Eq.~\eqref{eq:kurtz_delay}) that is correlated to a realisation of the nM-TLM approximation (Eq.~\eqref{eq:tau_leap_update}) through the use of the same Poisson process clocks.

Using the complete Poisson processes, we arrive at the coupled non-Markovian next reaction method (C-nM-NRM). The algorithm generates an exact realisation of a non-Markovian reaction network that is correlated to given $\tau$-leaping realisation as generated from Alg. 4. As discussed in the main text, the method presented here in Alg.~\ref{alg: nm-nrm-coupled} is effectively the nM-NRM (Alg. 3 in main text) that is driven by the completed unit-time Poisson processes constructed using Alg. 5 from the main text.  
\begin{algorithm}[H]
	\caption{Coupled non-Markovian Next Reaction Method (C-nM-NRM)}
	\begin{algorithmic}[l]
		\Require A non-Markovian biochemical reaction network with $\mathcal{N}$ species and $\mathcal{M}$ reaction channels, of which $\mathcal{D}$ are delay reactions (Eq. (6)); an initial time $t_0\leq0$; a final time $T > t_0$; an initial system state $\mathbf{x}_0=\mathbf{X}(t_0)$; event time sets $\mathcal{T}_j$ for $j \in \mathcal{R}$ and $\mathcal{T}_j^\kappa$ for $\kappa = 1,2, \ldots, K_j$ and $j \in \mathcal{R}_D$ as constructed from Alg.~5;
		\State Initialise state $\mathbf{X} \leftarrow \mathbf{x}_0$, time $t \leftarrow t_0$, internal times $t_j \leftarrow 0$ for $j \in \mathcal{R}$, first reaction intervals, $r_{j,u} \leftarrow \min{(\mathcal{T}_j)}$ for $j \in \mathcal{R}$, and delay sets $\mathcal{D}_j \leftarrow \emptyset$, $j \in \mathcal{R}_D$; and set event counters $m_j = 1$.
		\While{$t < T$}
		\State Update next reaction times $\tau_{j}$ for $j \in \mathcal{R}_I$, and $\tau^k_{j}$ for $k \in \mathcal{D}_j$ and $j \in \mathcal{R}_D$ (Eqs. (22)--(23));
		\State Find $\tau_{\mu_1} \leftarrow \min_{j \in \mathcal{R}}{(\tau_j)}$, and $\tau_{\mu_2}^k \leftarrow \min_{j \in \mathcal{R}_D, k \in \mathcal{D}_j}{(\tau_j^k)}$;
		\State Set next event time $\tau_{\text{min}} \leftarrow \min{(\tau_{\mu_1},\tau_{\mu_2}^k)}$
		\State Update remaining intervals $r_{j,u}$ for $j \in \mathcal{R}$, and $r_{j,u}^k$ for $k \in \mathcal{D}_j$ $j \in \mathcal{R}_D$ (Eqs (20)--(21));
		\State Update internal times $t_{j} \leftarrow t_{j} + \tau_{\text{min}}$ for $j\in \mathcal{R}$, and $t_{j}^k \leftarrow t_{j}^k + \tau_{\text{min}}$ for $k \in \mathcal{D}_j$ $j \in \mathcal{R}_D$;
		\If{$\tau_{\text{min}} = \tau_{\mu_1}$}
		\If{$m_j < |\mathcal{T}_j|$}
		\State Set $m_j \leftarrow m_j +1$;
		\State Set $t_{j,u}^{(m_j)} \leftarrow \min{(\mathcal{T}_j \backslash \{t_{j,u}^{(1)},t_{j,u}^{(2)},\ldots,t_{j,u}^{(m_j-1)} \})}$ and $t_{j,u}^{(m_j)} \leftarrow \min{(\mathcal{T}_j \backslash \{t_{j,u}^{(1)},t_{j,u}^{(2)},\ldots,t_{j,u}^{(m_j-2)} \})}$;
		\State Set $r_{\mu_1,u} \leftarrow t_{j,u}^{(m_j)} - t_{j,u}^{(m_j-1)}$;
		\Else
		\State Set $r_{\mu_1,u} \sim \text{Exp}(1)$;
		\EndIf
		\If{$\mu_1 \in \mathcal{R}_I$}
		\State Set $\nu \leftarrow \nu_{*,\mu_1}$;
		\Else
		\State Set $\nu \leftarrow \eta_{\mu_1} -\nu_{*,\mu_1}^{-}$;
		\State Set $t_{\mu_1,0}^{*,k'} \leftarrow t + \tau_{\mu_1}$, and $t_{\mu_1}^{k'} \leftarrow 0$; 
		\State Set $k' \leftarrow X^*_{\mu_1}+ 1$;
		\For{$\kappa \in [1,2,\ldots,K_j]$}
		\If{$\mathcal{T}_j^\kappa \neq \emptyset$}
		\State Select $t_{j,u}^{\kappa,(m)}$ from $\mathcal{T}_j^\kappa$ with probability $1/|\mathcal{T}_j^\kappa|$;  
		\State Set $r_{\mu_1,u}^{k'} \leftarrow t_{j,u}^{\kappa,(m)} - t_{\mu_1,0}^{*,k'}$ and $\mathcal{T}_j^\kappa \leftarrow \mathcal{T}_j^\kappa \backslash \{t_{j,u}^{\kappa,(m)}\}$;
		\State \textbf{Break loop};
		\EndIf
		\EndFor
		\If{$\kappa > K_j$} 
		\State Set $r_{\mu_1,u}^{k'} \sim \text{Exp}(1)$
		\EndIf
		\State Initiate delayed reaction $\mathcal{D}_{\mu_1} \leftarrow \mathcal{D}_{\mu_1} \cup \{k'\}$;
		\EndIf
		\Else
		\State Set $\nu \leftarrow \nu_{*,\mu_2}^+ -\eta_{\mu_2}$;
		\State Complete delayed reaction $\mathcal{D}_{\mu_2} \leftarrow \mathcal{D}_{\mu_2} \backslash \{k\}$;
		
		\EndIf
		\State Update the system state $\mathbf{X} \leftarrow \mathbf{X} + \nu$, and time $t \leftarrow t + \tau_{\text{min}}$;
		\EndWhile
	\end{algorithmic}
	\label{alg: nm-nrm-coupled}	
\end{algorithm}

\newpage
\section{Inter-event time distribution}
\label{app:interevent time}

\numberwithin{equation}{section}
\numberwithin{figure}{section}
\numberwithin{algorithm}{section}
\numberwithin{table}{section}
\setcounter{equation}{0}
\setcounter{algorithm}{0}
\setcounter{figure}{0}

To characterise the stochastic dynamics introduced by delay reactions, we begin by deriving the inter-event time distribution associated with their conclusion events. This leads naturally to a view of the system as a non-Markovian process, and motivates a generalisation of the reaction channel formalism central to stochastic simulation.

At time $t > t_0 > 0$ consider a delayed chemical species $\mathcal{X}^{*,k}_j$, created at initiation time $t_{j,0}^{*,k} < t$ and whose completion time is a random variable $T > t$. Suppose the reaction completes with propensity $\lambda^*_j(\mathbf{X}(t), t_j^k)$, where $\mathbf{X}(t)$ is the system state at time $t$ and $t_j^k=t-t^{*,k}_{j,0}$ is the internal time since the reaction initiation. Then, the conditional probability that the reaction will complete in the interval $[t, t+\Delta t)$ is given by,
\begin{align*}
	\condprob{T < t+\Delta t}{T > t} = \lambda^*_j(\mathbf{X}(t+\Delta t), t_j^k+\Delta t)\Delta t + \mathcal{O}\left(\Delta t^2\right).
\end{align*}
Let $g_k(\tau) = \condprob{T > t+\tau}{T > t}$ denote the survival probability after some time $\tau$. Then, over an interval $\tau + \Delta t$ we have,
\begin{align}
	g_k(\tau+\Delta\tau) = g_k(\tau)\left(1-\lambda^*_j\left(\mathbf{X}(t+\tau+\Delta t), t_j^k+\tau+\Delta t\right)\Delta t + \mathcal{O}\left(\Delta t^2\right) \right). \label{eqn: delay 3}
\end{align}
Rearranging Eq.~\eqref{eqn: delay 3} and taking the limit $\Delta t \to 0$ gives an ODE for the survival function,
\begin{align}
	\dydx{g_k(\tau)}{\tau} &= -g_k(\tau)\lambda^*_j(\mathbf{X}(t+\tau),t_j^k+\tau), \quad\quad g_k(0) = 1 \label{eqn: delay ODE}
\end{align}
Here, Eq.~\eqref{eqn: delay ODE} admits the solution,
\begin{align}
	g_k(\tau) &= \exp{-\int_{0}^{\tau} \lambda^*_j\left(\mathbf{X}(t+s), t_j^k+s \right) \rmd s}. & \label{eqn: tau comp CDF}
\end{align}
The solution to the survival function (Eq.~\eqref{eqn: tau comp CDF}) mirrors the arrival time distribution of an inhomogeneous Poisson process with a propensity depending on internal time $t_k$. This leads to the system as a whole being non-Markovian due to the internal time dependence on the survival time, and the dependence on the history since the internal time depends on the initiation time $t_0^{*,k}$.  We note that our framework is applicable for all valid propensity functions, that is, $\lambda(\mathbf{X},t) \ge 0$ for all $ X \in \mathbb{N}^{\mathcal{N}+\mathcal{D}}$ and $ t > 0$, and $\lim_{t\to\infty} \int_0^t \lambda(\mathbf{X},u)\,\text{d}u =\infty$. This admits a wide class of possible waiting time distributions including, uniform, heavy-tailed distributions such as half-Cauchy, and deterministic delays via piecewise-constructed hazard functions.

\end{appendices}
\end{document}